\documentclass{vgtc}                          % final (conference style)
\ifpdf%                                % if we use pdflatex
  \pdfoutput=1\relax                   % create PDFs from pdfLaTeX
  \usepackage{graphicx}                % allow us to embed graphics files
  \DeclareGraphicsExtensions{.pdf,.png,.jpg,.jpeg} % for pdflatex we expect .pdf, .png, or .jpg files
\else%                                 % else we use pure latex
  \usepackage{graphicx}                % allow us to embed graphics files
  \DeclareGraphicsExtensions{.eps}     % for pure latex we expect eps files
\fi%

\graphicspath{{figures/}{pictures/}{images/}{./}} % where to search for the images
\usepackage{array}
\usepackage{multirow}
 \usepackage{amsmath}
 \usepackage{paralist}
\usepackage{microtype}                 % use micro-typography (slightly more compact, better to read)
\PassOptionsToPackage{warn}{textcomp}  % to address font issues with \textrightarrow
\usepackage{textcomp}                  % use better special symbols
\usepackage{mathptmx}                  % use matching math font
\usepackage{times}                     % we use Times as the main font
\usepackage{cite}                      % needed to automatically sort the references
\usepackage{tabu}                      % only used for the table example
\usepackage{booktabs}                  % only used for the table example

\onlineid{8909}

\vgtccategory{Research}

\vgtcinsertpkg

\newcommand{\tool}{StoryLensEdu~}
\newcommand{\toole}{StoryLensEdu}
\newcommand{\etal}{{\it et al.\ }}
\newcommand{\etc}{{\it etc.}}
\newcommand{\ie}{{\it i.e.,\ }}
\newcommand{\stitle}[1]{\noindent{\bf #1}}

\definecolor{tablerowcolor}{rgb}{0.667,0.667,0.667 }
\definecolor{tablerowcolor2}{rgb}{0,0,0}
\definecolor{visual}{HTML}{e8efd9}
\definecolor{motion}{HTML}{fde7d5}
\definecolor{narrative}{HTML}{e2dce9}
\definecolor{audio}{HTML}{d6ebf2}
\definecolor{bluecrayola}{rgb}{0.12,0.46,1.0}

\newcommand{\revise}[1]{{\color{black} #1}}
\newcommand{\review}[1]{\textcolor{purple}{}}

\ifpdf%                                % if we use pdflatex
  \pdfoutput=1\relax                   % create PDFs from pdfLaTeX
  \usepackage{graphicx}                % allow us to embed graphics files
  \DeclareGraphicsExtensions{.pdf,.png,.jpg,.jpeg} % for pdflatex we expect .pdf, .png, or .jpg files
\else%                                 % else we use pure latex
  \usepackage{graphicx}                % allow us to embed graphics files
  \DeclareGraphicsExtensions{.eps}     % for pure latex we expect eps files
\fi%

\graphicspath{{figures/}{pictures/}{images/}{./}} % where to search for the images
\usepackage{array}
\usepackage{multirow}
 \usepackage{amsmath}
 \usepackage{paralist}
\usepackage{microtype}                 % use micro-typography (slightly more compact, better to read)
\PassOptionsToPackage{warn}{textcomp}  % to address font issues with \textrightarrow
\usepackage{textcomp}                  % use better special symbols
\usepackage{mathptmx}                  % use matching math font
\usepackage{times}                     % we use Times as the main font
\usepackage{cite}                      % needed to automatically sort the references
\usepackage{tabu}                      % only used for the table example
\usepackage{booktabs}                  % only used for the table example
\usepackage{algorithm}
\usepackage{algpseudocode}
\usepackage{amsmath} % 用于数学公式
\usepackage{float}

\title{
\textit{Supplementary Material for \toole}}

\title{\toole: Personalized Learning Report Generation through Narrative-Driven Multi-Agent Systems}

\author{
Leixian Shen\thanks{L. Shen and Y. Luo are co-first authors. E-mail: \{lshenaj, yluodj, rshengac, yhecy, haotian.li, lyangbb\}@connect.ust.hk, huamin@cse.ust.hk \\This paper has been accepted to IEEE PacificVis 2026 (Conference Track). The final published version will be available on the official website.}~, Yan Luo$^*$, Rui Sheng\thanks{R. Sheng is the corresponding author. }~, Yujia He, Haotian Li, Leni Yang, Huamin Qu \\
\scriptsize The Hong Kong University of Science and Technology
}

\abstract{
Personalized feedback plays an important role in self-regulated learning (SRL), helping students track progress and refine their strategies.
However, current common solutions, such as text-based reports or learning analytics dashboards, often suffer from poor interpretability, monotonous presentation, and limited explainability. 
To overcome these challenges, we present \toole, a narrative-driven multi-agent system that automatically generates intuitive, engaging, and interactive learning reports. \tool integrates three agents: a Data Analyst that extracts data insights based on a learning objective–centered structure, a Teacher that ensures educational relevance and offers actionable suggestions, and a Storyteller that organizes these insights using the Hero’s Journey narrative framework.  \tool supports post-generation interactive question answering to improve explainability and user engagement. We conducted a formative study in a real high school and iteratively developed \tool in collaboration with an E-learning team to inform our design. Evaluation with real users shows that \tool enhances engagement and promotes a deeper understanding of the learning process.

} % end of abstract

\CCScatlist{
  \CCScatTwelve{Human-centered computing}{Visu\-al\-iza\-tion}{}{}
}

\begin{document}


\firstsection{Introduction}

\maketitle

% \section{Introduction}

Self-regulated learning (SRL) is widely recognized as one of the most effective learning strategies, empowering students to take charge of their own learning by setting goals, monitoring progress, and reflecting on outcomes~\cite{Monique1999Self-regulated, Sheng2025}. Central to this process is personal feedback~\cite{Deborah1995Feedback}, which provides essential insights that help learners adjust their strategies and improve performance~\cite{VanMechelen2023}.
To support SRL, many online platforms such as Leetcode\footnote{https://leetcode.com/} and Quizizz\footnote{https://quizizz.com/}offer personalized feedback based on learners' practice and exam histories, helping them track progress and refine their approaches.
% \footnote{https://leetcode.com/}
% \footnote{https://quizizz.com/}

Given the pivotal role of feedback, recent research has increasingly focused on how to generate more effective student reports. The first direction is to enrich the content of these reports. Moving beyond traditional formats that only display grades or raw scores, modern reports now incorporate detailed analytics, such as practice frequency, time-on-task, accuracy~\cite{Susnjak2022}, and mastery of learning objectives~\cite{zhou2025study}, to provide a deeper understanding of student performance.
Though more insightful, the growing complexity of report content has outpaced the capabilities of traditional text-based reports. 
% and comprehensible
To address this, many systems now integrate visualizations to enable more engaging communication. For example, learning analytics (LA) dashboards~\cite{Fernandez-Nieto2024, Pozdniakov2023} is a commonly used format that organize student data into interactive, structured visuals. Susnjak \etal~\cite{Susnjak2022}, for instance, visualize weekly student activity to support exploration of learning patterns.
% ~\cite{Yan2024a, Fernandez-Nieto2024, Susnjak2022, Pozdniakov2023}

However, despite their strengths, existing solutions, especially LA dashboards or text-based reports face notable limitations. 
First, the sheer volume and complexity of the data can overwhelm students, making it difficult to interpret insights without proper guidance. Misinterpretations can also lead to misguided learning decisions~\cite{Fernandez-Nieto2024,Pozdniakov2023}. 
% ~\cite{Yan2024a,Fernandez-Nieto2024,Pozdniakov2023}
Second, dashboards often rely heavily on monotonous presentations (\ie direct list of charts and text), which may lack engagement, particularly for younger audiences like middle and high school students. This can result in overlooked opportunities for reflection and growth.
Third, these reports are typically one-way, requiring students to passively receive and interpret the feedback without the ability to ask follow-up questions for further explanation or guidance.

This research aims at a system for providing personalized feedback that is comprehensive, comprehensible, and engaging, and for providing question-answering assistants to personalize the feedback consumption process. To inform the system design, we first conducted a formative study in a practical high school setting to explore current practices and preferences around personalized feedback from both student and teacher perspectives. 
We derived a set of design considerations about the generation workflow, content requirements, and presentation format for the report.
% \leni{There should be some research findings that using stories enhanced learning results too, which can be cited to strengthen the motivation if you have time.}
% \haotian{Please add a few sentences to summarize the key findings that can motivate our design below.}
% \haotian{The tool name can be improved. Let's think more about it.}
Based on the findings, we developed \toole, a narrative-driven multi-agent system that automatically generates personalized, engaging, and interactive learning reports. 
The system was iteratively developed in collaboration with an actual E-learning platform team, including two key modules:
The first one is a \textit{report generation engine} with a multi-agent architecture.
The engine features a team of three LLM-powered agents with different roles. The \textit{Data Analyst} extracts key insights from student data based on a learning objective–centered structure. 
The generated insights are next evaluated by the \textit{Teacher} agent to ensure its educational value and offer customized suggestions.
Then, the \textit{Storyteller} organizes these insights and suggestions into a cohesive and engaging narrative with human-like storytelling techniques. 
Specifically, we embed the Hero’s Journey narrative framework~\cite{Farmer2019, Wei2024} into the Storyteller agent.
The design allows the storytelling agent to transform the report into a compelling story, enriched with text and visuals that mirror the learner’s journey across different stages.
To further support personalized and interactive exploration, we carefully designed an \textit{interaction module}.
It enables post-generation question answering based on user-selected visual or textual elements within the report.
Finally, we evaluate \tool with real-world users and demonstrate its effectiveness in enhancing engagement and delivering actionable insights.
% into student performance.

\revise{The main contribution is a multi-agent system architecture that automates context-aware visualization recommendation by dynamically coupling data insights with narrative-driven storytelling. 
The seamless coupling of automated insight extraction, narrative framing, and interactive Q\&A represents a meaningful advancement in how visualization systems can scaffold complex data for non-expert users.
% This framework establishes a seamless bridge between raw learning analytics and interactive visual Q\&A, providing non-expert learners with an interpretative scaffold to navigate complex personalized data.
}

\clearpage

\iffalse
Our contributions are as follows:

\begin{itemize}
    \item We conduct a formative study in a real high school to understand current feedback practices and expectations from both students and teachers.
    \item We design and develop \tool, a multi-agent system that automatically generates personalized, narrative-rich, and interactive learning reports. The system was iteratively developed in collaboration with an actual E-learning team.
    \item We evaluate the system with real-world users and show that it improves engagement and supports meaningful reflection on learning performance.
\end{itemize}

\fi
\section{Related Work}

\subsection{Student Performance Feedback}
% report and feedback 
While traditional text-based reports remain foundational for delivering summative and formative feedback to students, their dense textual format often results in cognitive overload, limiting their effectiveness in facilitating students' reflection~\cite{FernandezNieto2022}.

% LA dashboard
To address these limitations, Learning Analytics (LA) Dashboards have emerged as a promising solution. Susnjak et al.~\cite{Susnjak2022} developed an innovative dashboard employing machine learning algorithms to provide predictive insights, enabling students to interpret their learning progress and make informed adjustments. Additionally, Yan et al.~\cite{Yan2024a} demonstrated that integrating GenAI into LA dashboards enables more effective analysis of textual data, unlocking deeper insights into learning processes.
% \leni{I don't think Susnjak et al's work directly "generate actionable insights" (like telling the students you should do xxxx). The current expression can mislead me into thinking that learners only need to follow those insights and have no need to worry about understanding feedback.}

% Recent advancements enhanced explanatory design features by integrating visual elements to guide educators toward critical learning patterns \cite{Pozdniakov2023, Yan2024a}. \leni{for this work, you should explain it more concretely too.}

% Storytelling for Student Data
% 1. Narrative visualization 2. Data comics
Although LA dashboards mitigate textual overload through visual encoding, they are hard to interpret and fail to sustain engagement. Pozdniakov et al.~\cite{Pozdniakov2023} found that teachers can benefit a lot from explicit data storytelling guidance when interpreting data dashboards. Fernandez-Nieto et al.~\cite {Fernandez-Nieto2024} demonstrated how automated feedback can be transformed into data stories tailored to teachers' instructional goals, fostering reflective student practices. Some researchers also explore alternatives such as visualization, storytelling, and narrative structures to move beyond traditional dashboard paradigms~\cite{Echeverria2018, FernandezNieto2022}. Ruan et al.~\cite{math2020} showed that embedding analytics within narratives significantly enhanced children's engagement and conceptual understanding. 
% \leni{For the work of Milesi, be specific that they evaluated the effectiveness of data comics for...the current expression sounds like they achieved a system already, which made the research gap look incremental.} 
Furthermore, Milesi et al. evaluated the effectiveness of data comics in helping students visualize and reflect on personalized learning experiences ~\cite{Milesi2024}.
% \leni{Do we only advance previous work by adding a bidirectional analytical experience? It is contradictory to what I read in the introduction section.}
However, these narrative-driven reports are still constrained by a lack of interactive features ~\cite{Sun2024, Chaudhury2024} and fail to deliver a bidirectional analytical experience that enables users to delve deeper into the insights.
% Inspired by these approaches above, our work incorporates the Hero's Journey narrative structure to enhance our reports and designs an interactive post-generation question answering to enable students to freely explore more personalized data insights.

\revise{Existing feedback formats generally follow two paradigms: static summarization (text-based reports) and exploratory visualization (LA dashboards). 
While reports often lead to passive reception, dashboards require high data literacy to bridge the gap in interpretation. 
Our work introduces another paradigm, interactive automated narrative synthesis, combining automated synthesis with post-generation Q\&A. Unlike dashboards that prioritize data accessibility for manual exploration, our system prioritizes interpretative scaffolding. By shifting the burden of synthesis from the student to a multi-agent system, we transform fragmented metrics into a cohesive learning journey, moving beyond what the data is to what it means for the learner’s progress.}

\subsection{Data-Driven Article Generation}

Data-driven articles are a popular form of data storytelling that combine text, visualizations, and their interplay~\cite{Hao2024}. To streamline the traditionally manual process of writing narratives and crafting corresponding visualizations, researchers have proposed various interaction paradigms and automation techniques~\cite{li2024we}. One key direction supports tighter text-visual integration, through interactive manipulation~\cite{Latif2021}, real-time annotation~\cite{Masson2023}, markup languages~\cite{Conlen2021,Grimley2018}, programming interfaces~\cite{Horowitz2023}, and block-based editing~\cite{DataParticles2023, Chen2022a}.
With the rise of large language models (LLMs), recent work increasingly focuses on automating article generation~\cite{li2025reflection}. For example, DataTales~\cite{Sultanum2023} uses LLMs to generate textual narratives from charts; SNIL~\cite{Cheng2024} produces episode-based sports news from user input with support for post-editing; and DATAWEAVER~\cite{Fu2025} enables bidirectional composition between text and visualizations.

While these systems have significantly improved the learnability of data article creation, they often place less emphasis on enhancing expressiveness, especially in domain-specific contexts like education. In contrast, our work leverages LLMs to automate the data article generation pipeline while embedding narrative structures (e.g., the hero’s journey~\cite{Farmer2019, Wei2024, Mittenentzwei2023a}) to boost learner engagement and expand the expressive application in educational scenarios.

\subsection{Multi-Agent Systems for Data Storytelling}
Multi-agent systems have been increasingly adopted across various domains for their ability to coordinate specialized agents toward complex, goal-oriented tasks~\cite{Xi2023, lin2025survey, wang2025medkgi}.
These systems typically decompose workflows into modular roles, each handled by an autonomous agent with domain-specific capabilities.
\revise{Such paradigms have been shown to be more effective than single-agent approaches in handling complex scenarios that involve multiple subtasks~\cite{gao2025singleagentmultiagentsystemsboth}.}
For example, Data Director~\cite{datadirector} integrates agents representing a data perceiver, analyst, and designer to automatically generate data videos from tabular inputs. InsightLens~\cite{Weng2024} employs LLM-based agents to capture, organize, and visualize insights throughout the analysis process, supporting insight navigation through multi-faceted conversational contexts. 
LightVA~\cite{Zhao2024} introduces a planner–executor–controller loop for task planning and execution in visual analytics. 
LAVE~\cite{Wang2024} leverages LLM-powered agents to assist in video editing through language-based augmentation.

Building on this line of work, our system adopts a multi-agent architecture composed of a data analyst, storyteller, and teacher agent, designed to collaboratively generate personalized educational learning reports. \revise{This design reduces the risk of conflating these responsibilities and makes the trade-offs between data correctness, educational intent, and narrative quality more manageable.}

% \input{sections/formative}
% \section{System Overview}

\section{Learning Objective-centered Data Structure}
% \haotian{We should introduce why we need to define the data structure with a short paragraph here. }

% To enable personalized, objective-driven feedback, we define a structured representation of student data centered on learning objectives.  
% At the heart of the system is a learning objective-centered data structure, conceptualized as a directed graph of learning objectives.
% This enables consistent tracking of performance, facilitates alignment with curriculum goals, and allows downstream agents to extract, reason about, and present insights in a coherent way. 
% By structuring the learner’s knowledge state in terms of objectives, the system ensures that analysis and feedback are explicitly tied to clear criteria for success. This approach aligns with the idea that effective feedback should be referenced to explicit goals and standards rather than given in a vacuum~\cite{Nicol01042006}.
% In other words, the learning objective graph makes the criteria for “good performance” transparent: each node defines what the student should be able to do, and the attached performance data indicates how well the student is currently doing in relation to that goal. 

At the core of the system is a learning objective-centered data structure, modeled as a directed graph of interconnected learning objectives. This structure systematically tracks performance, aligns feedback with curriculum standards, and enables downstream agents to coherently extract, reason, and present learning insights. By explicitly linking a learner's knowledge state to defined objectives, the system ensures that feedback is directly referenced to transparent criteria, aligning closely with principles that effective feedback should be anchored to explicit goals rather than provided in isolation~\cite{Nicol01042006}.

\stitle{Learning Objective.}
A learning objective is a specific knowledge point that students should learn during an instructional unit.
We define a set of learning objectives that need to be mastered by a specific group of students as: $O = \{ \text{obj}_{1}, \text{obj}_{2}, \ldots, \text{obj}_{n} \}.$
% \[
% O = \{ \text{obj}_{1}, \text{obj}_{2}, \ldots, \text{obj}_{n} \}.
% \]
These learning objectives may exhibit dependencies among them, which can be represented as a directed graph \( G = (V, E) \), where \( V \) is the set of vertices, representing the learning objectives \( O \), and \( E \) is the set of directed edges, indicating the dependencies between the learning objectives.
% \begin{itemize}
%     \item \( V \) is the set of vertices, representing the learning objectives \( O \),
%     \item \( E \) is the set of directed edges, indicating the dependencies between learning objectives.
% \end{itemize}
For any two learning objectives \( \text{obj}_i \) and \( \text{obj}_j \) in \( O \), if there exists a directed edge from \( \text{obj}_i \) to \( \text{obj}_j \), which is denoted as \( (\text{obj}_i, \text{obj}_j) \in E \), 
% then \( \text{obj}_i \) serves as a prerequisite for mastering \( \text{obj}_j \).
then \( \text{obj}_i \) is referred to as a predecessor objective, indicating that \( \text{obj}_i \) serves as a prerequisite for mastering \( \text{obj}_j \).
\revise{The graph is automatically generated by parsing official curriculum standards. The system maps specific knowledge points as nodes and their predefined prerequisite relationships as directed edges.}

\begin{figure}[t]
\centering
\includegraphics[width=\linewidth]{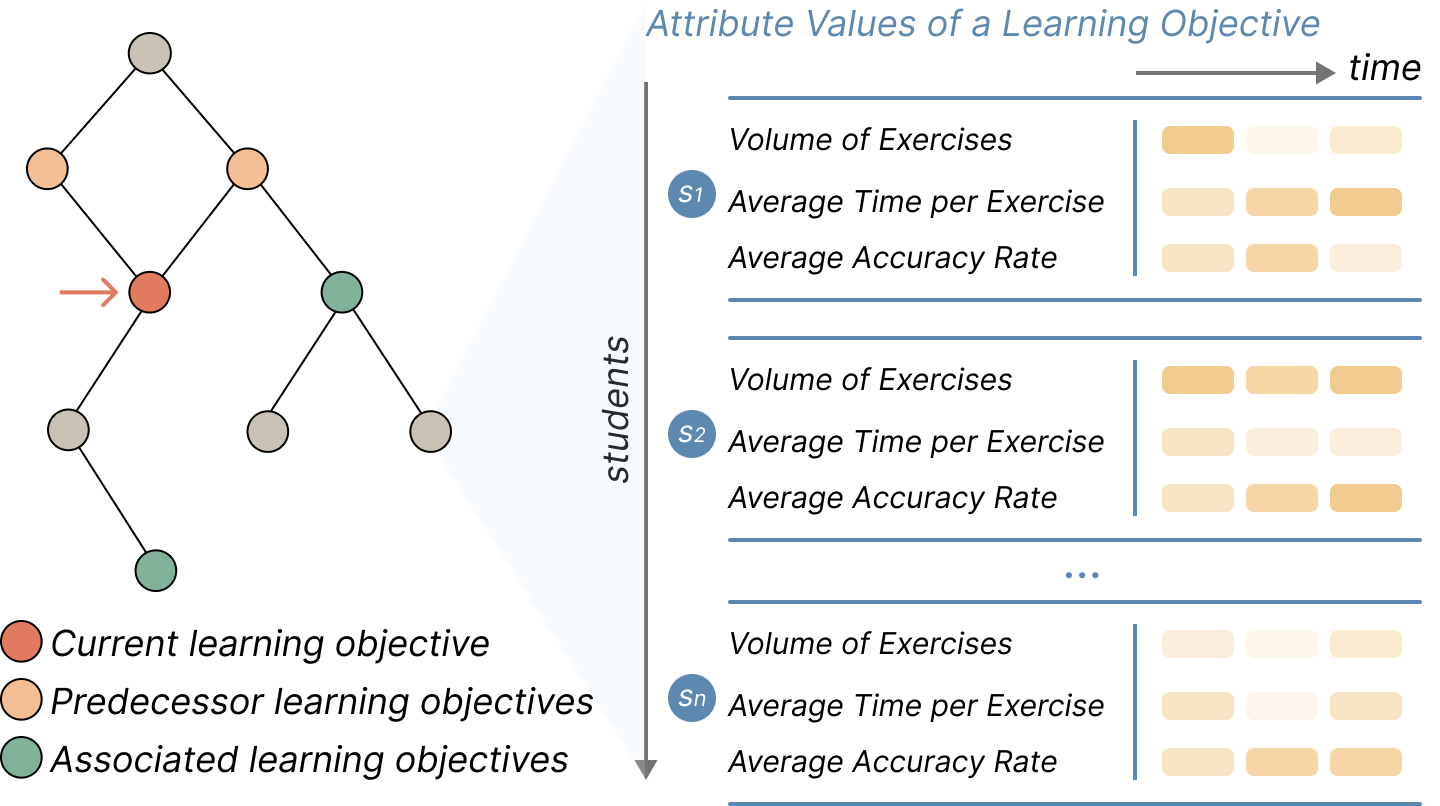}
\vspace{-5px}
\caption{Learning objective graph with three types of objectives.} 
\label{fig:LO}
\vspace{-10px}
\end{figure}

Based on the learning objective graph and the structural components of learning objectives within a student practice question, we define three types of learning objectives (Fig. \ref{fig:LO}): 
% current, predecessor, and associated nodes. These categories are essential for tailoring meaningful feedback and guiding review paths during personalized learning.
% These three types serve as the foundation for our algorithm to provide suggestions to students. 
% \haotian{We need to explain why these three types are important to consider, perhaps with one or two real-world cases. A few references would add more credibility to them. }
\textbf{(1) Current learning objective:} The learning objective that the user or system is currently focusing on;
\textbf{(2) Predecessor learning objectives:} The prerequisites for the current learning objective;
\textbf{(3) Associated learning objectives:} Other learning objectives are always considered alongside the current learning objective.
\iffalse
\begin{itemize}
    \item \textbf{Current learning objective.} The learning objective that the user or system is currently focusing on.
    \item \textbf{Predecessor learning objectives.} The prerequisites for the current learning objective.
    \item \textbf{Associated learning objectives.} Other learning objectives are always considered alongside the current learning objective.
\end{itemize}
\fi

Student errors are often interdependent. The three types capture these dependencies from distinct perspectives, enabling meaningful feedback and personalized review~\cite{adaptivekdd,xiachi2019}. 
% They help diagnose challenges, reveal prerequisite gaps, and suggest related topics to reinforce understanding.

% Student errors are often not isolated. The three types capture these dependencies from different perspectives. They are critical for generating meaningful feedback and informing personalized review paths, enabling the system to diagnose difficulties, identify prerequisite gaps, and recommend conceptually related topics to reinforce understanding~\cite{adaptivekdd,xiachi2019}.

\stitle{Attributes of Learning Objectives.}
Each student can be modeled as an instance of an abstract learning objective graph, where each node corresponds to a learning objective enriched with a set of attributes (e.g., volume of exercises, time spent, accuracy, \etc). The overall learning behaviors and outcomes of a student \( s \) on a specific learning objective \( \text{obj}_j \) are represented as a multi-dimensional time series \( \text{L}_j \), capturing the temporal dynamics of these attributes. Each node in the graph also encodes data from multiple students, enabling peer-based performance comparison.

% In this work, we define three important attributes for each learning objective \( \text{obj}_j \in O \): the volume of exercises attempted, the average time spent per exercise, and the average accuracy rate. The values of these three attributes can change over time for a specific student, ultimately represented by a multi-dimensional time series $\text{L}_j$.
% \haotian{It sounds too arbitrary. There can be many attributes for each objective. Then, why do we pick these three particularly? Some references or justification are essential.}
% We will introduce how to extract $\text{L}_j$ for each student based on our student question data.

\begin{figure*}[t]
\centering
\includegraphics[width=\linewidth]{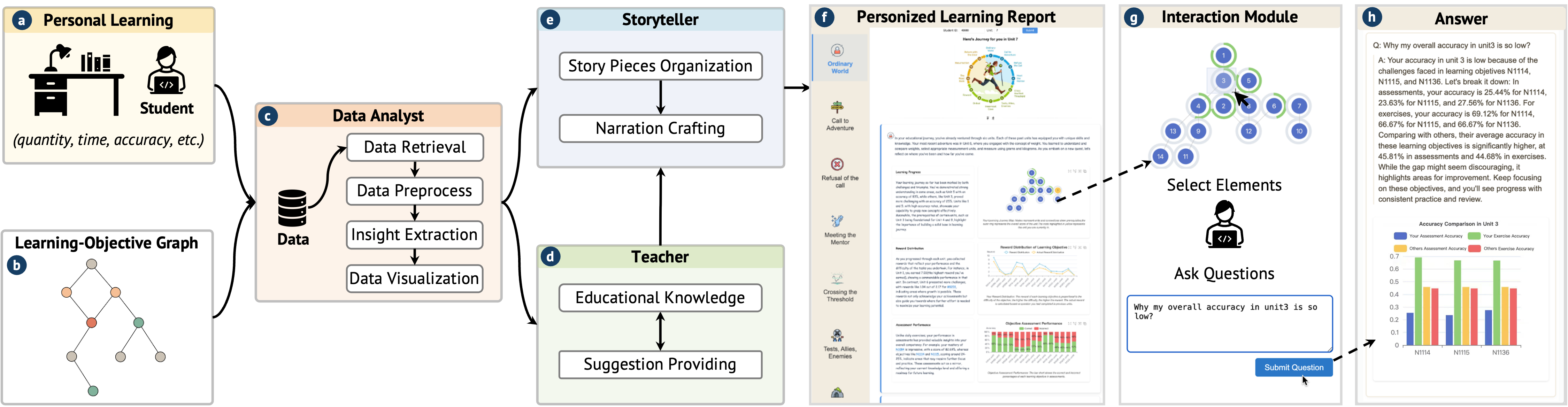}
\vspace{-5px}
\caption{
System Overview. The system takes student learning data (a) structured around a learning objective graph (b), which is analyzed by the Data Analyst agent (c) to generate diagnostic insights and visualizations. These are embedded into an annotated objective graph, then interpreted by the Teacher agent (d) to produce pedagogically grounded feedback. The Storyteller agent (e) organizes the content into narrative components using a Hero’s Journey structure and composes the personalized learning report (f). Users can interact with the report (g) to explore details and access contextual explanations (h).
% System overview. Student learning data (a) based on learning objective graph (b) is processed by the Data Analyst agent (c) to generate data insights and visualizations, leading to an annotated learning objective graph, which is processed by the Teacher agent to give professional suggestions. All the information is finally handled by the Storyteller agent (e) to organize story pieces based on the hero's journey narrative framework, and automatically generate the personalized learning report (f), enabling users for interactive exploration (g) to obtain further explanations (h).
} 
\label{fig: overview}
\vspace{-10px}
\end{figure*}

We next describe how to extract \( \text{L}_j \) for each student based on our example dataset in a real-world E-learning platform. While we focus on a core set of attributes in this paper, the approach is extensible to incorporate additional dimensions of learning behavior in future applications.

First, let \( S \) denote the set of a specific student group, and for each student \( s \in S \), we define a question set \( Q_s = \{ q_1, q_2, q_3, \ldots, q_i \} \), where \( q_i \) represents the \( i \)-th exercise attempted by the student.
Each exercise \( q_i \) is associated with a set of attributes \( A_i \) to characterize it. In our case,  \( A_i \) can be described as $A_i = \{ t_i, d_i, u_i, o_i, c_i \}$,
where
     \( t_i \) represents the absolute timestamp of when the exercise \( q_i \) was initiated by the student $s$;
     \( d_i \) represents the time duration taken to complete the exercise \( q_i \);
     \( u_i \) indicates whether the exercise was completed correctly; 
     % (binary outcome: correct or incorrect);
     \( o_i \subseteq O \) denotes a set of learning objectives associated with the exercise \( q_i \);
     \( c_i \) denotes the difficulty (challenge degree) of the question, \ie easy, medium, hard.

% \begin{itemize}
%     \item \( t_i \) represents the absolute timestamp of when the exercise \( q_i \) was initiated by the student $s$,
%     \item \( d_i \) represents the time duration taken to complete the exercise \( q_i \),
%     \item \( u_i \) indicates whether the exercise was completed correctly (binary outcome: correct or incorrect),
%     \item \( o_i \subseteq O \) denotes a set of learning objectives associated with the exercise \( q_i \).
%     \item \( c_i \) denotes the difficulty (challenge degree) of the question, \ie easy, medium, hard.
% \end{itemize}
% Therefore, we can gather information about the time when a student initiated an exercise, the time spent on it, whether the final answer was correct, and the associated learning objectives for that exercise.

Next, we can extract the values of the three attributes in a time interval \( T_k \) as follows:
\begin{compactitem}
    \item \textbf{Volume of Exercises}: The number of exercises \( N_j(T_k) \) attempted by the student \( s \) related to the objective \( \text{obj}_j \) during the time interval \( T_k \).
    \item \textbf{Average Time per Exercise}: The average time taken by student \( s \) to complete exercises with \( \text{obj}_j \) during the time period \( T_k \) can be calculated as:
    $$
    \bar{d}_j(T_k) = \frac{\sum_{i \in I_j(T_k)} d_i}{N_j(T_k)}
    $$
    % \[
    % \bar{d}_j(T_k) = \frac{\sum_{i \in I_j(T_k)} d_i}{N_j(T_k)},
    % \]
    where \( I_j(t_k) \) is the set of exercises related to \( \text{obj}_j \) attempted by student \( s \) during the time period \( T_k \).
    \item \textbf{Average Accuracy Rate}: The average correctness \( \bar{u}_j(T_k) \) for exercises associated with \( \text{obj}_j \) during the time period \( T_k \), calculated as:
    $$
    \bar{u}_j(T_k) = \frac{\sum_{i \in I_j(T_k)} u_i}{N_j(T_k)}
    $$
    where \( u_i \) indicates the outcome of each exercise (1 for correct, 0 for incorrect).
\end{compactitem}
Thus, the overall learning behaviors and outcomes of student \( s \) on a learning objective \( \text{obj}_j \) over time can be represented as a sequence:
$$
\text{L}_j(s) = \left\{ \left(N_j(T_k), \bar{d}_j(T_k), \bar{u}_j(T_k) \right) \right\}_{k=1}^{K}
$$

Given a student \( s \), this sequence encapsulates the evolution of learning behaviors and outcomes with a specific learning objective \( \text{obj}_j \) across the specified time intervals.

% \section{Report Generation Engine}
\section{\tool}
Based on the learning objective-centered data structure and the formative study insights (Please refer to supplementary materials for more details), we developed a multi-agent system, \toole, for report generation.
% , including three agent roles: data analyst, teacher, and storyteller.

\subsection{Overview}
% \haotian{Maybe we want to add one or two sentences to introduce why we need a multi-agent system and how the three roles are derived. Particularly, I wonder why data analysis and instructional feedback are finished by two agents.}
% To effectively generate personalized learning reports that are pedagogically meaningful, data-driven, and engaging, our system adopts a multi-agent architecture. 
This modular design separates complex educational tasks, such as data analysis, instructional reasoning, and narrative construction, into specialized roles.
The system overview is illustrated in Fig.~\ref{fig: overview}. 
A student’s self-regulated learning process continuously generates personal data (\textbf{a}), including exercise records, time logs, and performance across learning objectives, based on the learning objective graph (\textit{b}). 
Within the automatic \textit{Report Generation Engine} (\textbf{c}-\textbf{e}), 
the Data Analyst Agent (\textbf{c}) retrieves and structures student learning records based on student ID and learning unit. Leveraging the learning objective-centered graph, it extracts multi-dimensional performance sequences for each relevant learning objective, then mines key insights from these sequences across several dimensions (e.g., trend, anomaly, comparison). It also visualizes the insights to support interpretability.
% the Data Analyst Agent (\textbf{b}) retrieves and preprocesses relevant data based on the student ID and learning unit, transforming it into structured tables. It then automatically extracts performance insights and generates diverse visualizations. 
% These insights are passed to the Teacher Agent (\textbf{c}), which synthesizes them with predefined learning objectives to produce personalized instructional suggestions. 
These insights are then sent to the Teacher Agent (\textbf{d}), which contextualizes them within instructional goals. It selects and prioritizes feedback based on both the student’s current performance and learning objective dependencies (e.g., prerequisites and associations), and crafts pedagogically meaningful suggestions.
% , including motivational language and guidance on review paths.
Next, the Storyteller Agent (\textbf{e}) integrates both the analytical insights and instructional feedback into a coherent narrative using the Hero’s Journey structure. It assembles story elements aligned with the student’s progression and refines the tone and content to ensure the report is both informative and engaging.
The output is a personalized learning report (\textbf{f}), combining narrative text, performance insights, and rich visualizations. Students can further interactively explore the report via the \textit{Interaction Module} (\textbf{g}). They can select report segments or visual elements and ask questions, and the system will generate multimodal responses grounded in both the selected context and the underlying learning objective graph (\textbf{h}). 
% A complete example walkthrough can be found in supplementary materials.

% Both outputs are then forwarded to the Storyteller Agent (\textbf{d}), which organizes the report narrative using the Hero’s Journey framework. It composes coherent story elements, including insights and teacher suggestions, and crafts narrations aligned with the teacher’s tone and intent. 
% Finally, the engine outputs a personalized learning report (\textbf{e}). Students can explore the report and interact via the \textit{Interaction Module} (\textbf{f}), selecting specific text or visual elements to ask follow-up questions in natural language. The system responds with multimodal explanations grounded in the report context.

From an educational theory perspective, this unified, objective-driven workflow addresses all three key questions of effective feedback identified by Hattie and Timperley~\cite{Hattie}, \ie ``\textit{Where am I going?}'' (the learning objectives define the goals), ``\textit{How am I going?}'' (the Data Analyst and Teacher agents identify the student’s current performance status), and ``\textit{Where to next?}'' (the Teacher and Storyteller agents provide suggestions on next steps)''. 
It operationalizes formative assessment by not only diagnosing learning gaps but also immediately offering adaptive feedback~\cite{Fuchs1997}.

% It supports self-regulated learning by giving students clear goals and actionable feedback, which encourages them to reflect, set targets, and take steps to improve – in effect, helping them to “monitor and adjust” their learning strategies as self-regulating learners do. 
% And it follows cognitive apprenticeship principles by providing guided support (through the Teacher agent’s coaching) and then gradually empowering the student to take charge of their learning – the feedback gradually shifts from being very directive (when the student is novice in a topic) to more confirming and self-reflective as the student gains mastery, akin to the fading of scaffolding~\cite{collins1987cognitive}.

\subsection{Data Analyst Agent}\label{DataAnalyst}
The Data Analyst is primarily responsible for extracting insights from our organized student learning data to provide summative feedback and formative feedback.

% the Antv GetInsights\footnote{https://ava.antv.antgroup.com/en/api/insight/auto-insights} method, 

\subsubsection{Summative Feedback}\label{summative}
For summative feedback, we employ an automatic approach to mine insights from the constructed multivariate sequences \( \text{L}_j \). 
\revise{In our data formulation, Assessment Mode (Exercise or Test) and Learning Objective are treated as categorical dimensions alongside the temporal sequences. Consequently, the input to our algorithm is a multi-dimensional dataset where each record contains the performance metrics \( \text{L}_j \)
associated with specific dimension values. 

For summative feedback, our system processes these multi-dimensional datasets to identify significant learning patterns. First, the mining algorithm iterates through the subspaces defined by these dimensions. This allows us to extract data insights across diverse insight types (e.g., majority, outlier, trend, change point, and low variance) in different dimensions. Then the algorithm outputs raw insight tuples containing the insight type (e.g., Outlier), dimension (e.g., specific learning objective, accuracy), and the specific data evidence. These patterns offer students multifaceted views of their performance, such as identifying typical behaviors (majority), anomalies (outliers), or significant shifts (change points), which are critical for self-monitoring and reflection. 
}
% \[
% \text{L}_j(s) = \left\{ \left(N_j(t_k), \bar{t}_j(t_k), \bar{u}_j(t_k) \right) \right\}_{k=1}^{K}
% \] 
% We extract data insights across diverse analytical dimensions (e.g., majority, outlier, trend, change point, and low variance) to reveal interpretable learning patterns. These patterns offer students multifaceted views of their performance, such as identifying typical behaviors (majority), anomalies (outliers), or significant shifts (change points), which are critical for self-monitoring and reflection. 

To prioritize the most informative insights, we adopt the significance scoring framework proposed by Tang \etal\cite{topk}, which ranks insights based on a composite of statistical significance (p-value) and market share analysis~\cite{cooper1993market}. We surface the top three-ranked insights to help students efficiently recognize key performance signals and take informed, targeted actions. 
\revise{
Finally, to ensure precision, these findings are algorithmically pre-calculated as a quantitative ground truth rather than being generated by the LLM. Following a metadata-first workflow, the Teacher agent acts as a deterministic tool-user, retrieving this immutable structured JSON layer to generate interpretable natural language explanations (see Fig.~\ref{fig: report}(G)). 
% This architecture structurally prevents numerical hallucinations—a reliability validated by our user study, where experts gave high ratings for 'Data Insights Clarity' (4.5/5.0) without reporting factual errors.
% Finally, to make these technical findings interpretable and actionable for students, we feed these raw tuples into the Teacher agent. The Teacher agent generates the natural language explanations shown in the final report (see Figure 3(G)).
}
% Students can also use the interaction module (Sec.~\ref{sec:interaction}) for further insight mining.
% These insights support SRL by enabling students to detect learning gaps, track progress, and adjust strategies in a timely and data-driven manner.
% This method analyzes data across various dimensions (e.g., majority, outlier, trend, change point, and low variance) to extract meaningful patterns while prioritizing the insights based on their significance value. The method for calculating the significance of insights from multi-dimensional data is described in the paper by Tang et al.\cite{topk}, which uses a scoring function based on impact and significance. The impact measures the influence or market share~\cite{cooper1993market} of an insight and is calculated by summing the market share for all sibling groups and normalizing it. The significance of an insight reflects how unexpected the observed result is compared to what would be expected under a null hypothesis. It is measured using a p-value, and the method applies different hypotheses depending on the type of insight. The significance score increases as the p-value decreases, indicating a more surprising insight. We adopt the Top 3 insights as auxiliary aids to assist students in understanding their exercise and assessment performance from multiple dimensions.
% \haotian{Why are these types of data insights important? Can we explain their implications to the SRL scenario (maybe a table and examples)?}

\subsubsection{Formative Feedback}
For formative feedback, we first identify the learning objectives that need attention and suggestions.
Specifically, this agent continuously maps students' data onto the relevant learning objectives. 
% For each objective in the graph, the Data Analyst agent computes indicators of the student’s attainment; for example, it may estimate a mastery level (e.g., 85\% mastery of $Obj_i$), flag objectives where common errors occur, or calculate learning velocity (improvement rate) over time. 
% To derive an indicator that reflects the difficulty of a learning objective, the Data Analyst will first extract all questions from the database that contain $Obj_i$. Then, the Data Analyst will calculate the percentage of questions based on easy, medium, and hard categories relative to the total number of questions.
% The agent assigns weights of 1, 2, and 3 to the easy, medium, and hard questions, respectively. By calculating a weighted sum based on these percentages, we can obtain a value that represents the potential reward for mastering $Obj_i$. Consequently, the greater the difficulty of the objective, the higher the potential reward.
% In essence, the Data Analyst transforms low-level data into an assessment profile organized by objective, quantifying the learner’s progress (or difficulties) per objective~\cite{Outay2024}.
To estimate the difficulty and instructional value of a learning objective \( \text{Obj}_i \), the Data Analyst agent first retrieves all relevant questions tagged with \( \text{Obj}_i \) from the database. It then calculates the distribution of these questions across difficulty levels (easy, medium, and hard), expressed as proportions of the total. Each difficulty level is assigned a weight, 
% (e.g., 1 for easy, 2 for medium, and 3 for hard), 
and a weighted sum is computed to derive a scalar \textit{reward score} for \( \text{Obj}_i \). This score reflects the potential instructional gain or effort required to master the objective—higher values correspond to more challenging objectives.

After a student completes a learning unit, the system computes an actual reward by combining question difficulty with the student’s accuracy. In parallel, the Data Analyst calculates several additional indicators per objective in the learning graph, such as estimated mastery level (e.g., 85\% mastery of \( \text{Obj}_i \)), frequency of common errors, and learning velocity (\ie improvement over time). Collectively, these indicators form a structured diagnostic profile that quantifies student progress, aligns with curriculum goals, and supports objective-specific feedback~\cite{Outay2024}.

For each selected learning objective, we analyze it within the broader context of the learning objective graph. Unlike traditional data storytelling approaches~\cite{Wang2020i,datadirector}, which primarily rely on flat or hierarchical data table relationships, our method captures topological and semantic relationships among nodes, structured through the learning objective graph~\cite{remex}. These relationships, grounded in the three defined types (current, predecessor, and associated learning objectives), enable more meaningful insights and relevant feedback.
% enable more meaningful insight synthesis and pedagogically relevant feedback.
% \haotian{This part sounds particularly interesting. I think there has been no study to search and organize insights based on topological/semantical relationships (mostly are based on data or hierarchical relationships among data tables). Also, we may consider it as an example of meta relations: https://arxiv.org/pdf/2501.03603. We should carefully formulate this problem and show its value. }

\begin{compactitem}
    \item \textit{Current learning objective.} The learning outcomes of the current learning objective may be closely related to its practice process. So we consider the insights extracted for this node in the summative feedback as potential formative feedback.
    \item \textit{Predecessor learning objectives.} The reasons for insufficient learning of the current learning objective may be related to the need for improvement in its predecessor nodes. Therefore, we apply the same insight extraction approach from the multi-dimensional time series of its predecessor nodes. Notably, if a predecessor node has further predecessors, we continue exploring backward.
    \item \textit{Associated learning objectives.} Students may perform well on questions involving a single learning objective but may struggle when multiple learning objectives are associated. To determine whether suggestions should be provided based on associated learning objectives, we organize the multivariate sequences \( \text{L}_j \)
    for a associated objective set \( J = \{ j_1, j_2, j_3, \ldots, j_q \} \), where $J$ is summarized from the questions practiced by students. Specifically, we discard associated objective sets that include any predecessor nodes of the current learning objective, as we have already considered them separately.
    % Finally, we extract insights for ${L}_{J}(s)$. 
\end{compactitem}

% \haotian{Since mentioning peers, we may want to discuss the privacy and peer pressure issues in the discussion section.}
% To reduce the search space, we map the \rui{dimensions of the selected data insights (how to describe it??)} to the learning data of relevant peers, providing an aggregated result for reference and comparison, thereby enabling students to make informed decisions about their learning strategies.
% After extracting insights across the three defined learning objective types, the agent selects the most relevant ones to formulate targeted suggestions. 
\revise{The extraction method follows that described in Sec.~\ref{summative}.}
Additionally, the extracted insights across the three types are also contextualized through peer comparison to highlight relative performance. This processed information (the updated objective graph with performance annotations) is then passed along to the Teacher agent.
% \revise{To ensure factual precision, \tool separates data mining from narrative synthesis. The Data Analyst agent first generates a structured metadata layer (JSON) containing immutable performance facts. The Teacher and Storyteller agents are strictly grounded in this metadata, preventing the system from hallucinating scores or performance patterns not present in the logs.}
% The Data Analyst agent’s role is purely analytic; it does not decide how to convey feedback to the student. 
% The Teacher agent handles that pedagogical interpretation. However, by supplying an objective-by-objective analysis of performance, it lays the foundation for personalized, objective-focused feedback.
By operating at the granularity of learning objectives, 
% the Data Analyst agent provides a fine-grained diagnosis of the student’s learning state. 
the Data Analyst agent essentially answers the formative question: ``\textit{Where is the learner right now with respect to each learning goal?}'' This objective-centric diagnosis is crucial for effective formative feedback~\cite{Hattie, Fuchs1997}. 
Rather than giving a single overall score or generic remarks, the system can say, e.g., ``\textit{you have mastered Objectives A and C, is making progress on B, but is struggling with Objective D.}'' 
Such specificity not only empowers more actionable feedback downstream but also aligns with established principles in educational assessment, which emphasize the value of tying feedback directly to learning goals~\cite{Nicol01042006}.
% This level of detail allows subsequent agents to generate feedback that is precise and actionable. Notably, this design also mirrors best practices in educational measurement, where aligning assessment data with learning objectives improves the relevance of feedback and remediation strategies~\cite{Nicol01042006}.

\begin{table*}[t!]
\scriptsize
% \small
\centering
\caption{Adaptation of hero's journey narrative framework to personalized learning report generation.}
% \vspace{-7px}
\label{tab: Hero}
\renewcommand\arraystretch{1}
\setlength{\tabcolsep}{1mm}{
\begin{tabular}{m{1.2cm}m{4cm}|m{4.8cm}m{4.9cm}m{1.7cm}}
\hline
\multicolumn{2}{c}{ \textbf{Hero's Journey Narrative Framework}} & \multicolumn{3}{c}{\textbf{Adaptation to Personalized Learning Report Generation}} \\ \hline
\textbf{\textit{Phase}} & \textbf{\textit{Stage}} & \textbf{\textit{Mappings to Learning Stages}} & \textbf{\textit{Story Piece Description}} & \textbf{\textit{Information}} \\ \hline
\multirow{3}{*}{Departure} & S.1 Ordinary World & Review prior knowledge & Prior learning objectives and performance & \multirow{3}{*}{\shortstack[c]{Overview and\\Introduction} } \\ \cline{2-4}
 & S.2 Call to Adventure & Start a new learning unit & Introduction to this unit's objectives &  \\ \cline{2-4}
 & S.3 Refusal of the call & Anxiety about a new unit & Introduction to the new unit challenges &  \\ \cline{1-5}
\multirow{3}{*}{Initiation} & S.4 Meeting the Mentor & Learning through lectures or guidance & Emphasis on acquired knowledge &  \multirow{6}{*}{\shortstack[c]{Summary\\Information}}\\ \cline{2-4} 
 & S.5 Crossing the Threshold & Preparing for exercises & Introduction to upcoming unit exercises &  \\ \cline{2-4}
 & S.6 Tests, Allies, Enemies & Practicing with various problems & Insights based on exercise performance &  \\ \cline{1-4}
\multirow{3}{*}{Unification} & S.7 Approach to the innermost cave & Preparing for the test & Emphasis on accumulated practice &  \\ \cline{2-4}
 & S.8 The Ordeal & Taking the unit test & Exam performance with comparisons &  \\ \cline{2-4}
 & S.9 The Reward & Mastering basic unit knowledge & Summary of what has been learned &  \\ \hline
\multirow{3}{*}{Return} & S.10 The road back & Completing unit learning tasks & Learning task completed & \multirow{3}{*}{\shortstack[c]{Formative\\Guidance}}\\ \cline{2-4}
 & S.11 The resurrection & Receiving feedback and taking action & Suggestions based on performance &  \\ \cline{2-4}
 & S.12 Return with the Elixir & Ready to move on with full understanding & Hope for future improvement &  \\ \hline
\end{tabular}}
\vspace{-10px}
\end{table*}

\subsection{Teacher Agent}\label{teacher}

The Teacher agent receives the diagnostic performance map from the Data Analyst and creates pedagogically meaningful feedback for each learning objective. It leverages pedagogical rules, domain knowledge, and best practices to determine the appropriate guidance for the student. For each objective, the Teacher agent evaluates the student’s performance (mastery level, errors, attempts, \etc.) and generates tailored feedback. If the student has demonstrated strong mastery of an objective, the Teacher agent will respond with positive reinforcement and perhaps an extension suggestion (e.g., ``\textit{You have a solid understanding of Objective A. Great job! You might try a more challenging problem involving this concept to deepen your skills.}''). If the student is struggling with an objective, the Teacher agent will provide constructive, specific advice on how to improve (e.g., ``\textit{Objective B is still challenging for you. I noticed confusion in applying formula X – reviewing the underlying principle of Y and practicing those types of problems will help. Let’s focus on that, and remember the strategy we discussed…}''). In cases of partial understanding, the agent might both praise what is correct and coach the student on what is missing, following the ``medal and mission'' approach (highlighting successes and next steps)~\cite{Fawzi2020}.

% \haotian{I feel somehow lost about the connection between this paragraph and the previous one. }
% Importantly, the Teacher agent’s feedback is objective-aligned and criterion-referenced; it focuses on the learning task and the content of the objective, rather than on the student personally. This reflects a key tenet of effective feedback: it should be task-centered, not ego-centered~\cite{Black1998}.
By centering comments on how the student performed relative to the objective’s criteria, the agent ensures the feedback is informative (about the work) and not judgmental (about the person). 
The Teacher agent essentially acts as an automated tutor, emulating the role of a human teacher engaging in formative assessment. It uses the evidence of learning (from the Data Analyst) to adapt its instruction and guidance to the learner’s needs~\cite{Fuchs1997}. 
Research has shown that such formative feedback, when done well, can substantially boost learning gains~\cite{Fuchs1997}, especially for students who are struggling~\cite{Black1998}, by explicitly addressing misunderstandings and providing clear routes for improvement.
The design of the Teacher Agent also draws on the principles of cognitive apprenticeship~\cite{collins1987cognitive}, which emphasize modeling expert thinking, providing scaffolded guidance, and fostering learner independence. 
For example, the agent makes expert thinking visible, explaining not just what is right or wrong, but why (e.g., ``Revisiting concept Y will help you address challenges in Objective B'') based on the learning objective graph.

\begin{figure*}[h!]
\centering
\includegraphics[width=\linewidth]{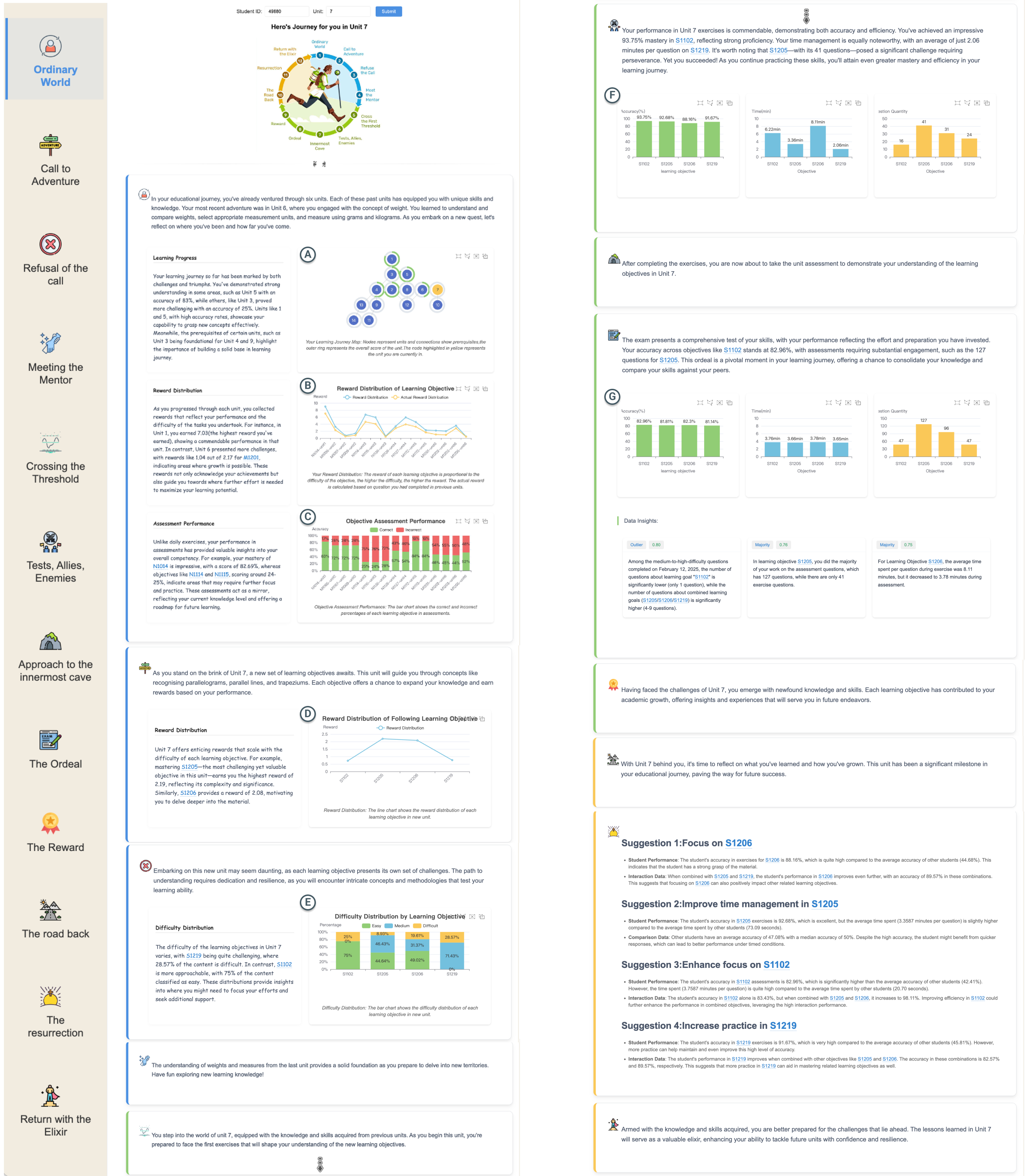}
\vspace{-10px}
\caption{
Report example. The left sidebar serves as an interactive index for rapid navigation, while the right pane displays detailed content.
% An example of generated report, including the index on the left and the content on the right that corresponds to the narrative structure of the hero's journey.
% The report consists of three sections: on the left is the table of contents, in the middle is the main content section, and on the right is a dialog panel. In this figure, both the left and right columns are part of the main content section and are connected seamlessly.
} 
\label{fig: report}
\vspace{-12px}
\end{figure*}

\begin{figure*}[t]
\centering
\includegraphics[width=\linewidth]{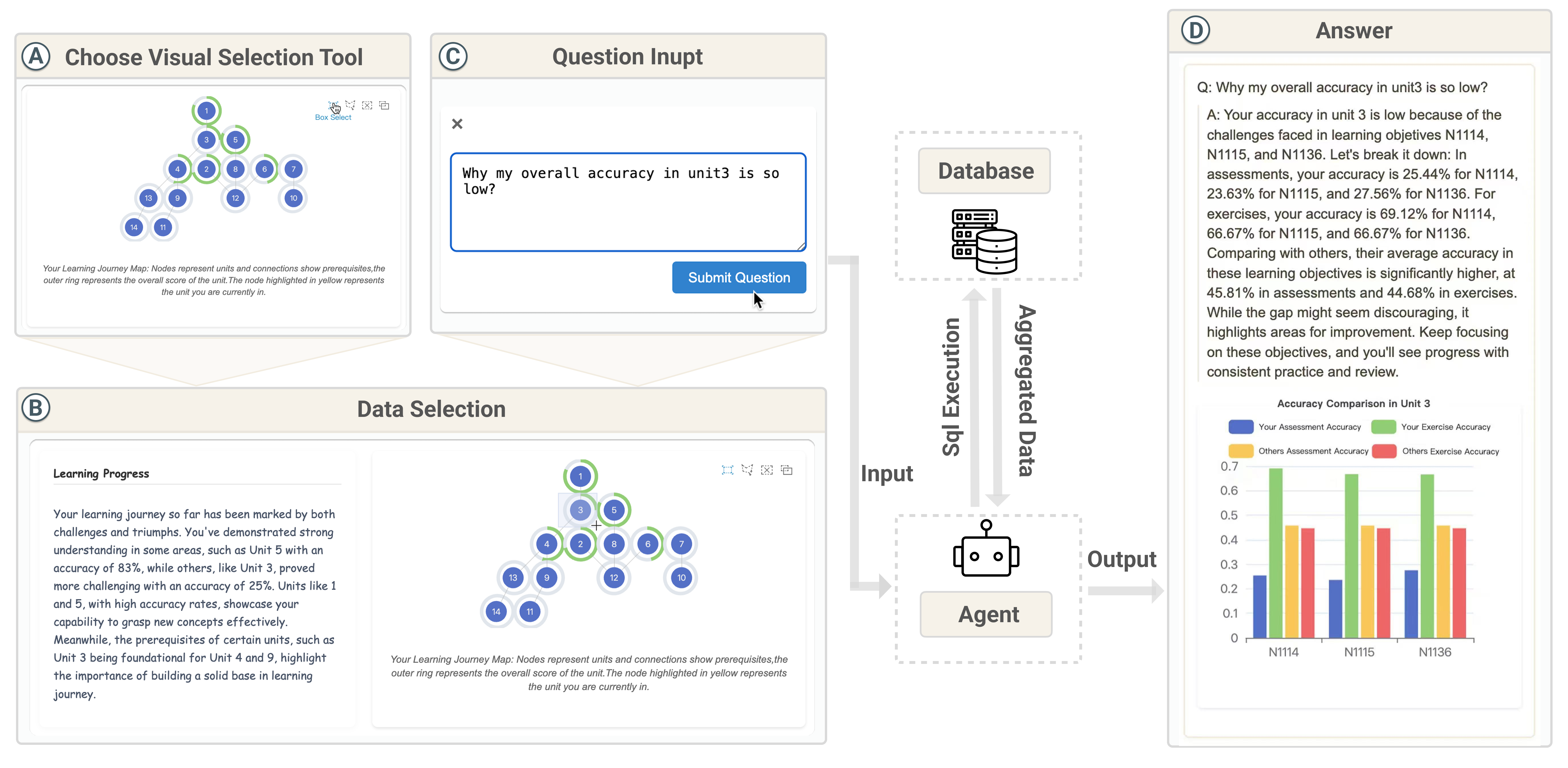}
\vspace{-12px}
\caption{
(A) Choose a visual selection tool (e.g., box or lasso). (B) Brush to choose the data points of interest. (C) Type the question and then submit it. (D) The agent retrieves related data from the database to generate the answer, and returns textual and visual elements.
% (A) Choose visual selection tool: Choose the box or lasso method to select the data point. (B) Data selection: Brush to choose the data point you're interested in. (C) Question Input: Type your question in this box and then submit it. (D) Answer: The agent receives the selected data point and question, retrieves related data from the database to generate the answer, and returns textual and visual elements to be presented in the interface.
} 
\label{fig: case1}
\vspace{-12px}
\end{figure*}

% \begin{figure*}[t]
% \centering
% \includegraphics[width=0.98\linewidth]{figures/case2.png}
% \vspace{-5px}
% \caption{Examples of interaction modules, where users select the text segments and ask further questions, and the agent automatically generate visualization-enriched answers.
% } 
% \label{fig: case2}
% \vspace{-10px}
% \end{figure*}

\subsection{Storyteller Agent}
The Storyteller agent is responsible for synthesizing the various objective-level feedback pieces from the Teacher agent and the insights extracted from the Data Analyst agent into a coherent, personalized feedback report that is clear and engaging for the student. While the Teacher agent focuses on what to say instructionally, the Storyteller agent focuses on how to say it in a narrative form that the learner will find accessible and motivating. 

% By presenting the feedback as a continuous narrative, the Storyteller agent enhances personalization and student engagement. 
A list of disjointed bullet points per objective can be factual but might not be very motivating or easy to digest. 
In contrast, a narrative format helps the student see the big picture of their learning journey: how the objectives they have mastered are accomplishments to be proud of, and how the challenges they face are part of a normal learning process with specific remedies available. Moreover, the Storyteller carefully calibrates the feedback tone to maintain the student’s motivation and self-efficacy. It balances praise and constructive critique so that the student remains confident to improve without feeling discouraged or complacent. 
% This approach is supported by research showing that students are more receptive to feedback that is supportive in tone, neither too harsh nor unrealistically positive, as they will more likely accept and act on such feedback.

Based on the performance insights and instructional suggestions generated by the Data Analyst and Teacher Agents, the Storyteller Agent employs the Hero’s Journey narrative framework to organize the report into a coherent and engaging story structure. 
\revise{The Hero’s Journey is a twelve-stage narrative archetype that has been widely used across educational and storytelling contexts to enhance emotional engagement and perceived relevance~\cite{Farmer2019, Wei2024, Mittenentzwei2023a}. 
% In the learning feedback setting, this structure provides a compelling metaphor for the personal learning journeys of the students: transforming them into protagonists navigating challenges, overcoming setbacks, and achieving growth.
We selected the Hero’s Journey over simpler structures, such as the three-act model, because its cyclical and transformative nature aligns more closely with the iterative process of self-regulated learning. While a three-act structure (setup, confrontation, resolution) captures linear progress, the Hero’s Journey emphasizes the internal transformation of the protagonist through setbacks and trials. This reframes learning not as a series of scores, but as a journey where ``Tests'' (S.6) and ``Ordeals'' (S.8) are essential, non-permanent stages of growth. Furthermore, the 12-stage framework provides the granularity necessary to map specific educational milestones, such as prior-knowledge review (S.1), formative practice (S.6), and post-exam reflection (S.11), to distinct narrative beats, ensuring the feedback is both cohesive and actionable.}

Specifically, we map each student's personalized learning process for a given unit to the twelve stages of the Hero’s Journey (see Tab.~\ref{tab: Hero}). Each stage corresponds to a distinct story piece, capturing elements such as initial goals (e.g., mastering learning objectives), encountered difficulties (e.g., exercises), turning points (e.g., exams), and reflections (e.g., teacher's suggestions). 
% The Storyteller Agent synthesizes narrative elements by aligning the structural needs of the Hero’s Journey with user-centered content considerations in our formative study, while seamlessly incorporating insights and suggestions from the prior two agents. 
\revise{To manage cognitive load, the twelve stages are hierarchically grouped into four macro-phases (Tab.~\ref{tab: Hero}), following an ``overview-first, detail-on-demand'' logic. Stages 1–3 provide a high-level review of goals, stages 4–9 focus on core performance analysis, and stages 10–12 deliver actionable guidance. 
This structure ensures students grasp the big picture before exploring specific diagnostic details. Furthermore, several stages serve as brief narrative transitions (e.g., S.4, S.10) to maintain story flow without adding analytical density.}
The result is a report that not only delivers informative feedback but also promotes reflection and engagement through a personalized, story-driven format. An report example is shown in Fig.~\ref{fig: report}.

% We have developed an interaction module for post-generation Q\&A that empowers students to explore more personalized and information-enriched insights. When students encounter questions about specific content or wish to delve deeper into details not fully addressed in the report presentations, they can freely select relevant text excerpts or visual charts to initiate interactive follow-up inquiries for extended clarification.
% When students freely explore within the report, they can use their mouse to select narrative segments of interest and ask questions. Similarly, they can brush or click on relevant data points in charts to delve deeper into the information. When handling large volumes of data or multiple learning objectives, the system will automatically present answers in the form of visual charts to enhance clarity and reduce cognitive overload for students. It is worth noting that we aim to address students' inquiries in a more supportive tone throughout the response process. In addition to providing them with more personalized insights tailored to their answers, we also offer emotional encouragement and support to empower their learning journey.

% \section{Post-Generation Q\&A}
\subsection{Post-Generation Interaction}\label{sec:interaction}
To support deeper understanding, we developed an interaction module for follow-up Q\&A with the generated report.

\subsubsection{Interaction Design}
Students can initiate contextual inquiries by selecting specific narrative segments or visual elements within the generated report. 
% For instance, they may highlight particular text passages or brush relevant data points within visualizations to formulate targeted questions. 
The system then generates detailed responses, enriched with visual explanations when necessary~\cite{IAI, Instructions}.
% , particularly for complex datasets or multiple interrelated learning objectives. 
% This interactive design reduces cognitive overload and enhances clarity by providing immediate and relevant insights. Responses are consistently delivered in a supportive and motivational tone, fostering student engagement and confidence throughout their learning journey.
Fig.~\ref{fig: case1} demonstrates this interactive questioning process. The student uses the visual selection tool to highlight specific data points of interest, such as a node representing Unit 3 in the learning progress graph, indicating weaker performance due to its smaller outer ring area. Subsequently, the student inputs a question (e.g., \textit{``Why is my overall accuracy in Unit 3 so low?''}) into the pop-up interface and submits it. 
% The selected data points and associated questions are then processed by the backend model.
Upon receiving this information, the backend identifies relevant learning objectives and performs an analysis grounded in the three learning objective types.
% : current, predecessor, and associated objectives. 
In the example, the system first identifies Unit 3 as encompassing three learning objectives (N1114, N1115, N1136), extracts related exercise and assessment performance data, and compares the student's metrics with peer averages to clarify potential areas for improvement. 
% As illustrated in Figure~\ref{fig: case2}, 
Students may also select combined textual and visual elements to generate queries, and the system dynamically determine and present the most suitable visualizations to effectively address their inquiries.

\subsubsection{Question Answering Model}
The Q\&A model is tightly integrated with the learning objective-centered data structure and its underlying knowledge graph to support precise, context-aware responses. When a student selects a text or visual element to pose a query, the system first uses the graph’s topology to map the selection to relevant learning objectives.

% The Q\&A model is tightly integrated with the learning objective-centered data structure and its underlying knowledge graph to support precise, context-aware responses. When a student selects a text or visual element to pose a query, the system maps the selection to relevant learning objectives using the graph’s topology, and then executes SQL queries against the structured learning database to aggregate relevant data and generates contextually meaningful visualizations.

Instead of executing complex, pre-defined analytical queries, the system then performs targeted SQL queries to retrieve a holistic dataset pertinent to these objectives. For single-objective queries, this dataset includes performance insights along three structured dimensions: the current objective, its predecessors, and its associated objectives. This tri-level analysis helps uncover both surface-level and latent learning difficulties. For multi-objective queries, data is retrieved across all selected objectives to identify higher-order patterns in interrelated learning behaviors. To enrich interpretability, peer comparison metrics are also incorporated into the dataset.

This curated dataset, combined with the student's natural language query and a predefined prompt template, is dynamically engineered into a detailed prompt. The LLM then processes this prompt, performing an in-context analysis to formulate a natural language answer while simultaneously generating the Echarts configuration for an appropriate visualization. This LLM-centric architecture bypasses the rigidity of traditional query systems, allowing for a more intuitive and responsive analytical dialogue.

\revise{We utilize an asynchronous pipeline where student metrics are pre-calculated and cached. This separates heavy data aggregation from interaction, ensuring that real-time report generation is limited only by LLM inference time (\~5s) rather than database queries.}

\section{Usage Scenario}\label{sec:case}
In this section, we will use fictional data 
% \rui{ethical concern if we use high school student data}
% 我感觉为了真实性，可以写成我们用了某个学生在看报告说的话在这里用一个虚假的case展现出来，不然后面我乱想象steven会说啥有点不好
to go through the report generated automatically by \toole.
Assume a student named Steven, who just finished studying Unit 7 of the third-grade curriculum. After completing all the practice and exam questions, he hoped to see his learning progress. Therefore, he generated a report with \toole, as shown in Fig.~\ref{fig: report}.

When Steven first started reading the report, which said, \textit{``In this journey, you've already ventured ...''}, he felt as if he had entered an adventure world.
\revise{Through the learning journey map (\autoref{fig: report}-A), he realized that he had mastered many units well, such as Unit 1, Unit 2, and Unit 5, which made him feel very proud.}
However, he also acknowledged some shortcomings, such as in Unit 3, where the report showed he might have only scored 25\%. Through more detailed line charts and bar charts (\autoref{fig: report}-B and C), he learned that concepts N1114 and N1115 in Unit 3 were areas he had not mastered well, scoring around 24\% to 25\%.
Steven silently resolved to solidify his understanding of Unit 3.

\revise{After that, Steven started to carefully examine the assessment related to Unit 7.
Through the reward distribution visualization (\autoref{fig: report}-D), he discovered that Unit 7 had four concepts he needed to master.} Among them, S1205 and S2106 appeared to be the most challenging.
\revise{He also saw the difficulty distribution (\autoref{fig: report}-E). For example, about 49.02\% of the questions related to S1206 were rated as difficult.
This allowed him to recognize how challenging this unit is.}
%This indeed made him feel a bit anxious. He recalled how he had felt when he first started learning Unit 7. He always asked himself, \textit{``Will I be able to overcome these challenges?''}
% \textit{``It is possible!''} Steven recalled how he once felt a surge of confidence. He did not know why he had been so confident at that time, but now, having seen the report, he was clear. \textit{``I have done well in the previous unit, laying a solid foundation for my learning in this unit!''} This was also the reason he had been determined to start to study Unit 7. \textit{``The difficulties I mentioned before are really nothing''}, Steven thought.
% The first step was to do more practice questions. He remembered the days he had devoted himself to solving problems.
% These were the days of battling enemies. 
At the same time, the report's hints reminded Steven of those days when he faced tough questions every day. \textit{``Have I really defeated the enemies?''}
\revise{Upon examining the data, he found a 93.75\% mastery rate in S1102 (\autoref{fig: report}-F). Each question appeared to take an average of about two minutes. He felt a quiet sense of satisfaction.
However, he also noticed that he did not seem to have mastered S1205 well, which would require attention in the future.}

% But that was not a problem; the positive feedback from earlier made sure that this small issue with S1205 would not bring him down!

He continued to read, \textit{``If those practice questions were the enemies, then what I face next could be the boss.''} 
\revise{When he examined the exam-related assessments (\autoref{fig: report}-G), the accuracy rates for all four learning objectives exceeded 80\%, indicating solid overall mastery.}
The report also showed him various information regarding his time spent on questions, and its encouraging tone helped sustain his passion. 
\textit{``I still need to improve.''} He noted some insights mentioned further down the report, such as not having enough difficult questions in S1102.
The report mentioned that Steven had done quite well in mastering the concepts of Unit 7. \textit{``There is no reward more important than this recognition,''} Steven stated.

Finally, it was essential to focus on reflection and feedback. \textit{``The report showed me ways to continue improving, such as suggestions for S1206.''}
Through this report, Steven felt as if he had re-experienced his learning journey in Unit 7. He could see himself at the beginning, feeling lost and afraid of difficulties, then pushing forward, overcoming challenges... This immersive report guided him through his experiences in Unit 7 and helped him digest and absorb essential learning summaries and suggestions.

% \section{Evaluation}

% evaluate intuitive (personal data insights extraction and presentation), engaging (story), and interactive (interaction) learning reports

% analysis by \leixian{yan}, written by \leixian{leixian}

% \subsection{Participants and Procedure}
% 1. directly walk through the report

% 2. free-exploration (interaction)

% 3. interview + questionnaire

% questionnaire:
% % name, age, 学历，sex, whether using personal learning platform (if yes, how long)

% % Q: 

% \subsection{Findings}

% \input{tables/results}
\begin{figure*}[t]
\centering
\includegraphics[width=0.95\linewidth]{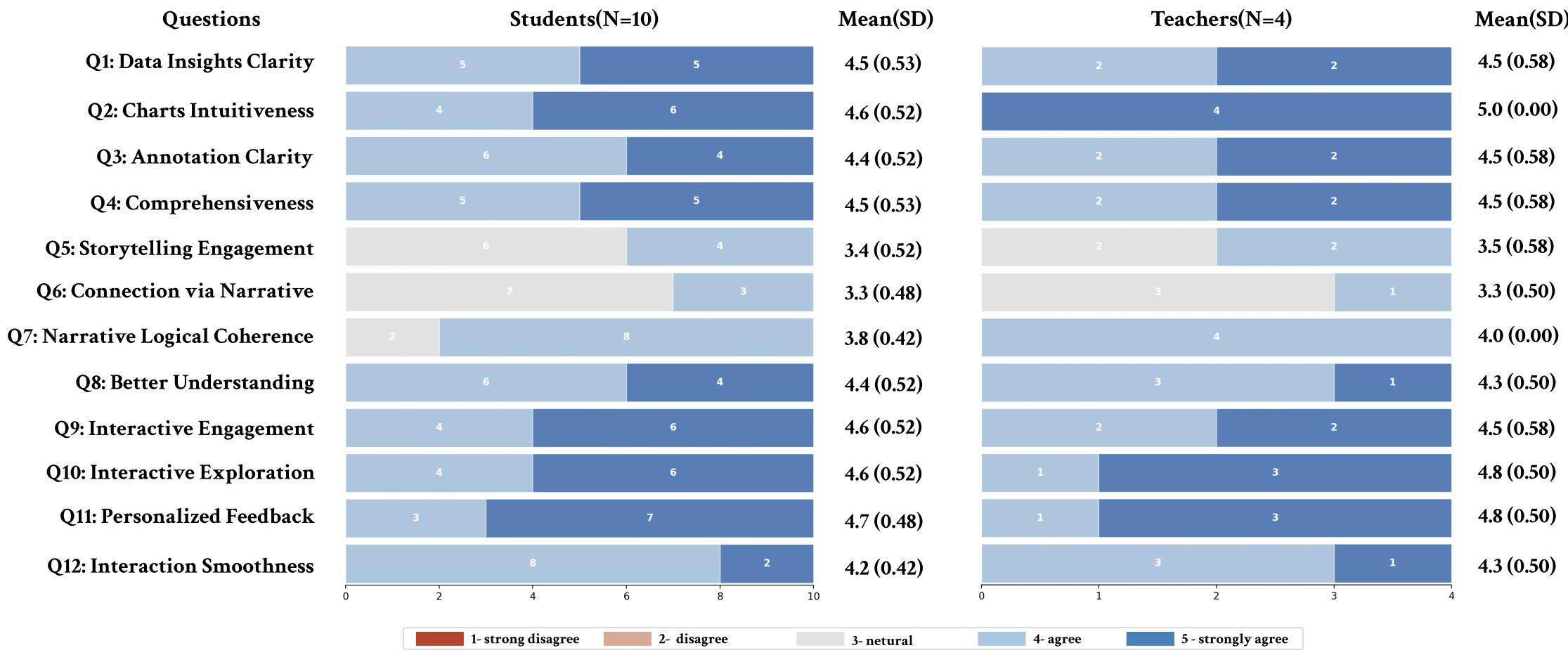}
\vspace{-5px}
\caption{Questionnaire results in the user study for 10 students (S1 - S10) and 4 teachers (T1 - T4).
} 
\label{fig: result}
\vspace{-10px}
\end{figure*}

\section{User Study}

Our evaluation aimed to assess the intuitiveness, engagement, and interactivity of the automatically generated personalized learning reports produced by \toole. 

\subsection{Participants and Procedure}
% We invited the same 10 students (S1 - S10) and 4 teachers (T1 - T4) who participated in our formative study to ensure consistency in evaluating the improvements implemented based on earlier insights. 
We invited 10 students (S1 - S10) and 4 teachers (T1 - T4) from a high school. 
Each student received their personalized report automatically generated by \tool based on real-world data, which includes their own learning data over two semesters and relevant data from over 1,000 peers.
Students were asked to read their reports carefully, employing a think-aloud method to articulate their thought processes.
% and initial reactions during their first interaction with the report. 
During their reading, students were asked to freely interact with the Q\&A module, exploring further details and raising inquiries as needed. 
Teachers randomly explored all students' reports.
Finally, we carried out individual semi-structured interviews to gain deeper insights into their experiences and asked each participant to complete a questionnaire.
% Finally, we conducted individual semi-structured interviews to delve deeper into their experiences and asked them to complete a questionnaire.
% , covering aspects of intuitiveness, engagement, and interactivity. 
The protocol has been approved by the IRB in our institution. 
\revise{All learning data is de-identified before transmission to LLM APIs. Personal identifiers (e.g., names, student IDs) are replaced with anonymized tokens.}
% , ensuring that the system processes performance patterns without accessing sensitive personal information.

% \subsection{Results}

\subsection{Quantitative Results}
The results from the questionnaire, as shown in Fig.~\ref{fig: result}, indicated generally positive feedback among both students and teachers. Specifically, clarity of data insights (mean=4.5, SD=0.53 for students; mean=4.5, SD=0.58 for teachers), intuitiveness of charts (mean=4.6, SD=0.52 for students; mean=5.0, SD=0.00 for teachers), and interactive exploration (mean=4.6, SD=0.52 for students; mean=4.8, SD=0.50 for teachers) were particularly praised, underscoring the system's effectiveness in visualizing complex educational data intuitively and interactively.
However, relatively lower ratings were observed in narrative engagement (students mean=3.4, SD=0.52; teachers mean=3.5, SD=0.58) and the connection of storytelling with personal learning behaviors (students mean=3.3, SD=0.48; teachers mean=3.3, SD=0.50). This indicates an opportunity to improve the narrative design to more closely resonate with the diverse expectations of the learners.

\subsection{Qualitative Findings}
Semi-structured interviews revealed rich insights into how participants perceived and interacted with \toole. A key theme among students was the sense of agency enabled by interactive elements. All students reported that features such as embedded Q\&A modules and annotated visualizations not only clarified their performance gaps but also empowered them to take action. As S4 noted, \textit{``The interactive visuals made it easy to pinpoint exactly where I struggled, and asking questions directly through the interface was really empowering.''} Students consistently described interactivity as pedagogically meaningful, supporting self-reflection and deeper engagement.
Teachers similarly valued visuals, particularly for diagnostic purposes. T2 remarked, \textit{``Charts simplified complex data clearly, allowing us to quickly grasp students' progress and intervene when necessary.''} Two teachers (T3, T4) highlighted that visualizations facilitated early intervention and cohort-level comparison.

The narrative component was broadly appreciated for structuring feedback into a coherent and accessible format. Participants described it as reducing intimidation and enhancing interpretability. T4 commented, \textit{``The narrative helps frame feedback in a way that feels more like a conversation than a judgment.''} S9 noted, \textit{``The report is enjoyable and relatable. I could better connect my learning actions with the outcomes shown.''} Many (10/14) endorsed the Hero’s Journey structure as logically coherent and pedagogically effective in connecting past performance, current outcomes, and future goals.

In addition, several participants proposed enhancements to improve personal resonance and emotional alignment. S7 suggested adapting the narrative tone and framing to better reflect individual learning preferences. While most students favored supportive language, a few preferred more provocative tones to enhance motivation. Thus, a uniform narrative style may limit engagement, and adaptive emotional framing could enhance relevance.
Teachers echoed the need for customization but cautioned against sacrificing specificity for fluency. As T1 noted, \textit{``The narrative sounds nice, but sometimes it glosses over the harder truths. We still need space for precise, critical feedback.''} 
% This highlights the challenge of balancing readability with pedagogical rigor, particularly in high-stakes contexts.
Participants (S1, S6, S7, T1) also raised concerns about information density, reporting difficulty in locating high-priority content. Suggested improvements included search, filtering, and visual anchors. T1 further proposed using animation to enhance immersion and data interpretability. Others (S2, S10) recommended incorporating imagery or comic-style illustrations to strengthen engagement.
Finally, T3 emphasized the need for narrative formats tailored to diverse educational levels and subjects, noting that, \textit{``The storytelling format suitable for senior-level math may not resonate with a lower-grade language class.''}
% (Please refer to supplementary materials for more discussions about the tool. )
% These insights underscore the importance of flexible, context-aware narrative strategies that accommodate both cognitive and emotional learner differences.

% \noindent
\revise{
\textbf{Failure Case Analysis.}
% While the multi-agent grounding mechanism significantly mitigates factual hallucinations, 
% We also found some failure cases during the user study. 
% While \tool received high ratings for clarity and intuitiveness, the scores for narrative engagement (Q5) and connection (Q6) were comparatively moderate (3.3–3.5). 
While \tool scored high in clarity, ratings for narrative engagement (Q5) and connection (Q6) were moderate (3.3–3.5). Qualitative interviews suggest this is not a rejection of the narrative paradigm, but rather a reflection of the ``one-size-fits-all'' nature of the current Hero's Journey framework. 
First, we observed narrative redundancy in reports for students with sparse activity data; because the 12-stage framework is currently rigid, the storyteller occasionally repeated the same insights to fill all sections. 
Second, there was a mismatch in tone, where the ``Heroic'' persona remained overly encouraging even when a student’s accuracy dropped significantly (e.g., calling a student a ``brave adventurer'' despite a 20\% success rate), which can gloss over critical gaps. 
Third, a prioritization gap occurred between statistical and pedagogical value; for instance, the system might highlight a rapid misclick as an insight into speed, which is educationally irrelevant. 
Fourth, the system sometimes exhibits logic leaps in pedagogical reasoning by misinterpreting conceptual associations as causal prerequisites. For example, finding a topological link between ``Functions'' and ``Geometry'' might lead the system to conclude that a student's struggle in ``Functions'' is due to poor ``Geometry'' skills, while ignoring the actual learning obstacle.
Finally, granularity issues in the learning objective graph can lead to vague advice, such as suggesting a general topic review instead of a specific procedural correction. 
}

\section{Discussion}

% \stitle{Personalization.}
% 1. more fine-grained, such as based on students' emotions, preference (some want supportive words, some need provocative language)

% 2. More detailed needs of different students (primary school students, college students) and different knowledge and skills (English, mathematics, programming)

% \stitle{Involve more Stakeholders.}
% involving more roles (Teachers, Parents, and Students)
% \cite{Han2024}

\stitle{Multi-Agent Collaboration.}
Our system introduces a multi-agent architecture where specialized agents emulate distinct education functions, mirroring human educator roles. This decomposition brings three key advantages. 
First, it enables functional specialization: each agent can be optimized for a narrow communicative goal, improving clarity and coherence. 
Second, it supports extensibility: new agents (e.g., peer tutor, parent mediator, or self-reflection companion) can be integrated without overhauling existing components. 
Third, it opens pathways for explainability. Users can trace the origin of each narrative element to a responsible agent and its rationale.
This architecture also introduces opportunities for intra-agent dialogue and negotiation. For instance, a tutor agent may prioritize conceptual feedback, while a motivational agent might advocate for positive framing. Designing protocols for agent-level conflict resolution or consensus-building could help balance feedback tone and focus in a pedagogically grounded manner~\cite{Xi2023}. Additionally, agent collaboration may support adaptive feedback pacing: agents could coordinate to stagger guidance over time, fostering sustained engagement rather than information overload. 

\stitle{Personalization.}  
This work demonstrates how student learning reports can be automatically generated with engaging narratives grounded in performance data. 
Moving forward, personalization can be deepened along several dimensions. One direction is emotional and motivational adaptation: students exhibit varied preferences for narrative tone; some favor encouraging language, while others respond better to more provocative feedback. Future systems could leverage sentiment analysis or learner profiling to dynamically adjust tone and emotional framing~\cite{Xie2023, Lan2023}. 
Participants also emphasized the importance of learner diversity across educational levels and domains. The structure, language, and visual density of feedback may require adaptation for different age groups (e.g., primary vs. college) and disciplines (e.g., mathematics vs. programming). Such contextual sensitivity can enhance clarity, relevance, and cognitive accessibility.
Personalization should also account for the involvement of additional stakeholders, such as parents or advisors, to support a more holistic feedback ecosystem~\cite{Han2024}. This could be enabled through multi-view reports or shared progress dashboards.
Personalization can evolve temporally as students’ goals, behaviors, and competencies change. This calls for longitudinal learner modeling and feedback strategies that evolve alongside the learner.
\revise{Additionally, a current limitation is the heavy reliance on numeric logs, which might overlook specific procedural errors. However, \toole’s multi-agent architecture is well-positioned to integrate LLM-based analysis of non-numeric artifacts. 
Future work could leverage a User Profiler agent to identify student preferences or learning styles. 
By feeding students’ verbal explanations, LaTeX-formatted derivations, or even OCR-processed sketches into the Data Analyst agent, the system could perform qualitative error pattern recognition. Future iterations will explore linking specific distractor analysis (common wrong answers) to the Storyteller agent, enabling even more nuanced narratives like, ``\textit{You consistently confused the sine and cosine rules, a common hurdle that we can overcome by...}''}

\stitle{Generalizability.}
% While \tool was developed and evaluated in a high school setting, its architecture is designed with transferability in mind. 
\revise{While evaluated in high school math, which is a prototypical testbed for validating the learning objective graph due to its explicit prerequisites, \tool is inherently domain-agnostic. Adapting to other STEM subjects requires only updating metadata and pedagogical rules, while for less hierarchical domains like the Humanities, the system can pivot from mastery-based to theme-based narrative logic.}
The system’s modular feedback pipeline, composed of objective-linked performance analysis, narrative construction, and role-specific agent contributions, can scale to contexts such as higher education (e.g., lab reports) and workplace training (e.g., compliance), and can also be extended to support other narrative formats~\cite{dataplayer, Zhang2025a, Gao2025, NarrativePlayer, NotePlayer}. Such flexibility stems from the structural encoding of objectives and the use of interpretable intermediate representations (e.g., knowledge graphs, reflection arcs), enabling customization without altering core logic. However, effective generalization requires retuning narrative style, granularity, and agent voice to align with the specific cognitive and emotional norms of each target domain~\cite{dataplaywright, DVSurvey}.
% can scale to diverse learning contexts. 
% In higher education, for instance, the same framework could be adapted to generate lab report feedback or project critique summaries. 
% In workplace training scenarios, narrative modules might focus on skill progression or compliance gaps.
% Crucially, the generalizability lies not only in technical modularity, but in the abstraction of educational roles and goals. 
% Because learning objectives are encoded structurally, and agents operate over interpretable intermediate representations (e.g., knowledge graphs, reflection arcs), the system can accommodate domain-specific customization without altering its core logic. However, effective generalization requires careful retuning of narrative style, feedback granularity, and agent voice to match the cognitive expectations and emotional norms of each target domain. 
% Future deployments should therefore pair technical adaptation with participatory design to ensure contextual alignment.

\stitle{Limitations.}
The system also presents several limitations. First, it relies on structured, objective-aligned performance data, which may not generalize to domains with unstructured or open-ended tasks. Extending the approach to such contexts will require more sophisticated data interpretation and knowledge extraction methods.
\revise{Second, our study prioritized evaluating the user experience of the proposed narrative paradigm. Future controlled trials will be conducted to quantify its impact on learning outcomes relative to established baselines, such as traditional dashboards and text summarization.}
% the current personalization is limited to observable performance metrics and does not account for learner affect, motivation, or long-term behavioral patterns. As feedback effectiveness is closely tied to emotional context, future work should incorporate richer learner models to enable more adaptive tone, pacing, and framing.
Finally, the system does not support direct teacher input during feedback generation, limiting opportunities for co-authorship. Enabling educator intervention could improve trust, contextual relevance, and classroom integration.

\section{Conclusion}

We presented \textit{\toole}, a narrative-driven, multi-agent system for generating personalized learning reports that are structured, intuitive, engaging, and interactive. 
% By combining learning objective-aligned analysis, visual storytelling, and interactive exploration, the system supports learners in understanding their progress and planning future actions.
Beyond automation, we argue that educational feedback should not only inform, but also motivate and guide. To that end, this work opens promising directions for collaborative human-AI feedback systems that evolve with learners, integrate diverse stakeholder perspectives, and adapt across domains and learning contexts.

\acknowledgments{
The authors would like to thank Trumptech for their continuous support. This work was supported in part by HK RGC GRF Grant 16218724 and HK ITF Grant PRP/017/22FX.}

\bibliographystyle{abbrv-doi}
\bibliography{main}

\appendix
\clearpage
\section{Formative Study}

We conducted a formative study to investigate current practices, challenges, and preferences regarding personalized feedback from both student and teacher perspectives, informing the design of an automated feedback generation system.

\subsection{Participants and Procedure} 
In a real-world high-school setting, we conducted semi-structured interviews with 10 students (6 males, 4 females; ages 16-18, S1–S10) and 4 teachers (2 males, 2 females; ages 22-36, T1–T4). Participants were selected based on their prior experiences with receiving and providing personalized feedback and their recognition of its importance in educational contexts.

Participants first described existing feedback practices, including typical content, formats, and perceived limitations. Next, they discussed ideal feedback they wished to receive or provide, explaining its potential impact on future learning. We explicitly probed about uncommon but valuable types of feedback and reasons for their scarcity. Additionally, we explored participants’ preferences for automatically-generated feedback reports, focusing on narrative and storytelling formats. Finally, participants sketched conceptual designs of personalized feedback reports, outlining sections, content specifics, and desired interactive features.

\subsection{Findings}
We present key findings from the interviews, enriched by direct quotes and insights, structured around motivation, challenges, content, and format preferences. These findings directly inform our hierarchical design considerations.

\subsubsection{Motivation and Challenges}
Both teachers and students emphasized personalized feedback as critical for effective learning. Teachers recognized its pedagogical value but described individual feedback as demanding substantial effort, resulting in superficial or delayed responses. Teacher T3 noted, \textit{“Giving personalized feedback takes hours, especially when we try to cover all learning objectives comprehensively. It’s easy to become generic or overly critical due to fatigue.”} Students underscored the motivational impact, with S2 remarking, \textit{“Constructive feedback makes me want to improve, while overly harsh or vague comments discourage me from engaging further.”} Thus, balancing specificity, timeliness, and encouragement emerged as a central challenge.

\subsubsection{Content Preferences}
Participants preferred personalized feedback closely aligned with explicit learning objectives, highlighting precise knowledge gaps and actionable next steps. Specifically, participants emphasized feedback should:

\begin{itemize}
    \item Provide a concise \textbf{performance overview}, followed by detailed insights into specific learning objectives.
    \item Identify key \textbf{errors or misconceptions}, recommending targeted future actions.
    \item Clearly \textbf{track progress over time}, enabling comparisons with personal historical performance and peer benchmarks.
    \item Suggest explicit, achievable \textbf{learning targets} and actionable steps.
    \item Maintain a supportive and encouraging tone, avoiding overly critical or vague language.
\end{itemize}

Students particularly valued feedback that combined summative overviews and formative guidance. S8 stated, \textit{“I want clear comparisons to my previous scores, but also specific advice on exactly how to improve next.”} They also appreciated direct links to additional resources for targeted review.

Notably, teachers identified valuable but rarely provided feedback types such as personalized emotional support or motivational messaging due to difficulties in objectively assessing student emotional states and preferences in tone.

\subsubsection{Format Preferences}
Participants consistently expressed strong preferences for visually enriched and interactive formats. Students highlighted that annotated visualizations significantly enhanced their understanding of performance data, making complex information more approachable. S5 commented, \textit{“Charts or diagrams help me visualize my progress much better than pure text, especially when combined with annotations that clarify what’s important.”} Teachers supported this view, noting visuals facilitated quicker diagnostics and intervention planning.

Interactive features were particularly valued by students, who appreciated being able to query specific aspects of their feedback dynamically. S3 remarked, \textit{“Interactive exploration lets me delve deeper into points I don’t initially understand, making the report much more useful.”}

Participants also indicated narrative-driven structures could enhance readability and emotional resonance. Teacher T4 emphasized the narrative’s potential, stating, \textit{“A well-organized narrative feels less judgmental and more like an ongoing dialogue with the student.”} Students similarly valued narrative coherence but stressed the need for personalized emotional and motivational framing, underscoring the limitation of generic storytelling formats.

\subsection{Design Considerations}
Drawing on these nuanced findings, we propose the following key considerations to effectively guide the design of automated personalized feedback systems:

\noindent
\textbf{D1. Automate Report Generation.}  
Given teachers’ emphasis on the intensive effort required for personalized feedback, automation is essential. Systems must integrate analysis, synthesis, and writing into seamless workflows, substantially reducing teacher workload while preserving feedback quality.

\noindent
\textbf{D2. Present Information in an Engaging and Accessible Format.}  
Feedback must be engaging, accessible, and coherent to motivate students to actively use it.

\begin{itemize}
    \item \textit{D2.1. Narrative Integration}: Employ structured storytelling to contextualize feedback meaningfully, bridging abstract data and personal learning journeys.
    \item \textit{D2.2. Interactive Exploration}: Enable interactive querying, filtering, or annotation to allow students to actively engage with and better comprehend their performance details.
\end{itemize}

\noindent
\textbf{D3. Ensure Personalized Feedback.}  
Feedback effectiveness significantly increases with personalization based on historical performance, peer comparisons, and individual emotional preferences.

\begin{itemize}
    \item \textit{D3.1. Individualized Analysis}: Leverage students’ historical data to identify precise learning trajectories and recommend targeted improvements.
    \item \textit{D3.2. Peer-based Contextualization}: Provide comparative insights that motivate improvement through realistic, actionable goals relative to peers.
\end{itemize}

\noindent
\textbf{D4. Align Feedback with Learning Objectives.}  
Explicitly align feedback with curricular objectives to ensure instructional relevance and facilitate direct pedagogical actions.

\begin{itemize}
    \item \textit{D4.1. Objective-driven Analysis}: Evaluate performance explicitly against defined curriculum objectives.
    \item \textit{D4.2. Guided Recommendations}: Provide actionable suggestions for review and improvement that align closely with identified objectives.
\end{itemize}

\noindent
\textbf{D5. Balance Summative and Formative Feedback with Supportive Tone.}  
Feedback should integrate both summary and formative components, delivered in a consistently constructive and motivating tone to sustain learner engagement.

\begin{itemize}
    \item \textit{D5.1. Summative Overviews}: Clearly summarize performance, track progress, and visually represent achievements to motivate continued improvement.
    \item \textit{D5.2. Formative Suggestions}: Offer specific, actionable steps tailored to students’ immediate learning needs.
    \item \textit{D5.3. Adaptive Emotional Tone}: Personalize the emotional framing of narrative content to align with diverse student motivational styles and emotional preferences, ranging from supportive to constructively provocative, as appropriate.
\end{itemize}

Incorporating these considerations ensures the proposed feedback generation system effectively addresses real-world challenges identified in our formative study, optimizing both pedagogical value and learner engagement.

\section{System Architecture}
The system architecture is rigorously designed to achieve an optimal balance between computational efficiency in data processing and the latency-sensitive requirements of personalized narrative generation. To ensure scalability and responsiveness, the backend functionality is decoupled into two synchronized modules.

\subsection{Backend Design}
The backend infrastructure comprises the following core components:

\subsubsection{Offline Data Processing Module}
Serving as the analytical backbone, this module operates asynchronously to handle heavy computational tasks. It interfaces with the MongoDB persistence layer to extract raw interaction logs, encompassing exercise attempts, assessment outcomes, and behavioral timestamps. A dedicated aggregation pipeline processes these logs to compute multi-dimensional performance metrics, such as accuracy, average response time, and completion volume, at the granularity of individual learning objectives. To mitigate the latency overhead during user interactions, these derived metrics are serialized into structured JSON objects and stored in a high-performance local cache, effectively isolating the real-time request loop from intensive database queries.

\subsubsection{Real-time Narrative Generation Module} Upon a user request, the system retrieves the pre-calculated metrics from the local cache. The Prompt Engine then integrates these metrics with the Hero's Journey narrative templates to construct a context-rich prompt. Finally, the LLM generates the personalized feedback story based on this prompt.

\subsection{Pseudo-code}

\begin{algorithm}[H]
\caption{StoryLensEdu Report Generation and Interaction Process}
\label{alg:storylens}
\begin{algorithmic}[1] % [1] 表示显示行号
\Require Student ID ($s$), Learning Unit ($U$), Learning Objective Graph ($G$)
\Ensure Personalized Report ($R$), Interactive Response ($Ans$)

\Statex % 空行用于分隔

\Procedure{StoryLensEdu\_Engine}{$s, U, G$}
    \State \textbf{Phase 1: Data Analyst Agent}
    \State \textit{\# Retrieve student's data for all learning objectives}
    \State $L_j \gets \Call{AggregateStudentData}{s, U, G}$ 
    \State $raw\_insights \gets \Call{ExtractTopK}{L_j, \text{significance\_framework}}$
    
   \State \textit{\# Analyze from three perspectives: current, predecessor, and associated objectives}
    \State $insights \gets \Call{TriLevelAnalysis}{raw\_insights, G}$
    
    \Statex
    \State \textbf{Phase 2: Teacher Agent}
    \State $pedagogical\_feedback \gets \emptyset$
    \For{\textbf{each} $insight$ \textbf{in} $insights$}
        \State $feedback \gets \Call{TeacherReasoning}{insight, \text{rules}}$
        \State $pedagogical\_feedback.\Call{Append}{feedback}$
    \EndFor
    \Statex
    \State \textbf{Phase 3: Storyteller Agent}
    \State \textit{\# Map the feedback into Hero's Journey framework}
    \State $structure \gets \Call{MapToHeroJourney}{ L_j,pedagogical\_feedback}$
    \State $R \gets \Call{GenerateReport}{structure}$
    \State \Return $R$
\EndProcedure

\Statex % 空行用于分隔两个过程

\Procedure{Interactive\_QA\_Module}{$R, query, selected\_context, G$}
    \State $target\_objs \gets \Call{MapToGraph}{selected\_context, G}$
  \State \textit{\# $data_{int}$: multi-objective interaction data}
\State \textit{\# $data_{cmp}$: other students' comparison data}
    \State $data_{int}, data_{cmp} \gets \Call{QueryDatabase}{target\_objs}$
    \State $Ans \gets \Call{LLMGenMultimodal}{query, data_{int}, data_{cmp}}$
    \State \Return $Ans$
\EndProcedure

\end{algorithmic}
\end{algorithm}

% \begin{figure*}[t]
% \centering

% \includegraphics[width=0.98\linewidth]{figures/case2.png}
% \vspace{-5px}
% \caption{Examples of interaction modules, where users select the text segments and ask further questions, and the agent automatically generate visualization-enriched answers.
% } 
% \label{fig: case2}
% \vspace{-10px}
% \end{figure*}

\subsection{Post-Generation Interaction}

To support a deeper understanding of the personalized report, we developed an interaction module that enables students to engage in a bidirectional analytical dialogue.

\subsubsection{Interaction Mechanism}
Students can initiate contextual inquiries by using a visual selection tool (e.g., a lasso or brush) to highlight specific data points, such as a node in the learning progress graph, or by selecting relevant narrative segments. Once the context is selected, the user inputs a natural language question (e.g., "Why is my overall accuracy in Unit 3 so low?") into the pop-up interface. This query, along with the selected context, is then transmitted to the backend for semantic parsing and data retrieval

\subsubsection{Context-Aware Backend Analysis}
The system identifies the specific learning unit or learning objectives associated with the user’s selection. Grounded in the learning objective graph's topology, the backend retrieves detailed data from three perspectives:
current objectives, predecessor objectives, and associated objectives. Additionally, if a user's query intention includes comparisons with others, the system will also retrieve corresponding information from those other users.

\subsubsection{LLM-Centric Visual Synthesis}
To ensure the response is both informative and intuitive, we employ an LLM-centric architecture that bypasses the rigidity of traditional query systems. The curated dataset and user query are fed into a predefined prompt template. Notably, this prompt instructs the LLM to perform an in-context analysis and return a dual-format response:

\begin{itemize}
\item \textbf{Natural Language Explanation:} A pedagogically grounded answer that interprets the data insights.

\item \textbf{Dynamic Visualization Code:} The LLM generates specific Echarts configurations (e.g., bar charts for comparisons or line charts for trends) tailored to the retrieved data.
\end{itemize}

The frontend then directly renders these Echarts configurations, providing the student with a multimodal response that combines narrative guidance with rich, interactive visualizations.

% \subsubsection{Interaction Examples}
\section{More Examples}
We provide three sample reports as references, representing students with varying academic performance and learning paces. These examples reflect the diversity of the generated outputs.

\begin{figure*}[t]
\centering
\includegraphics[width=0.98\linewidth]{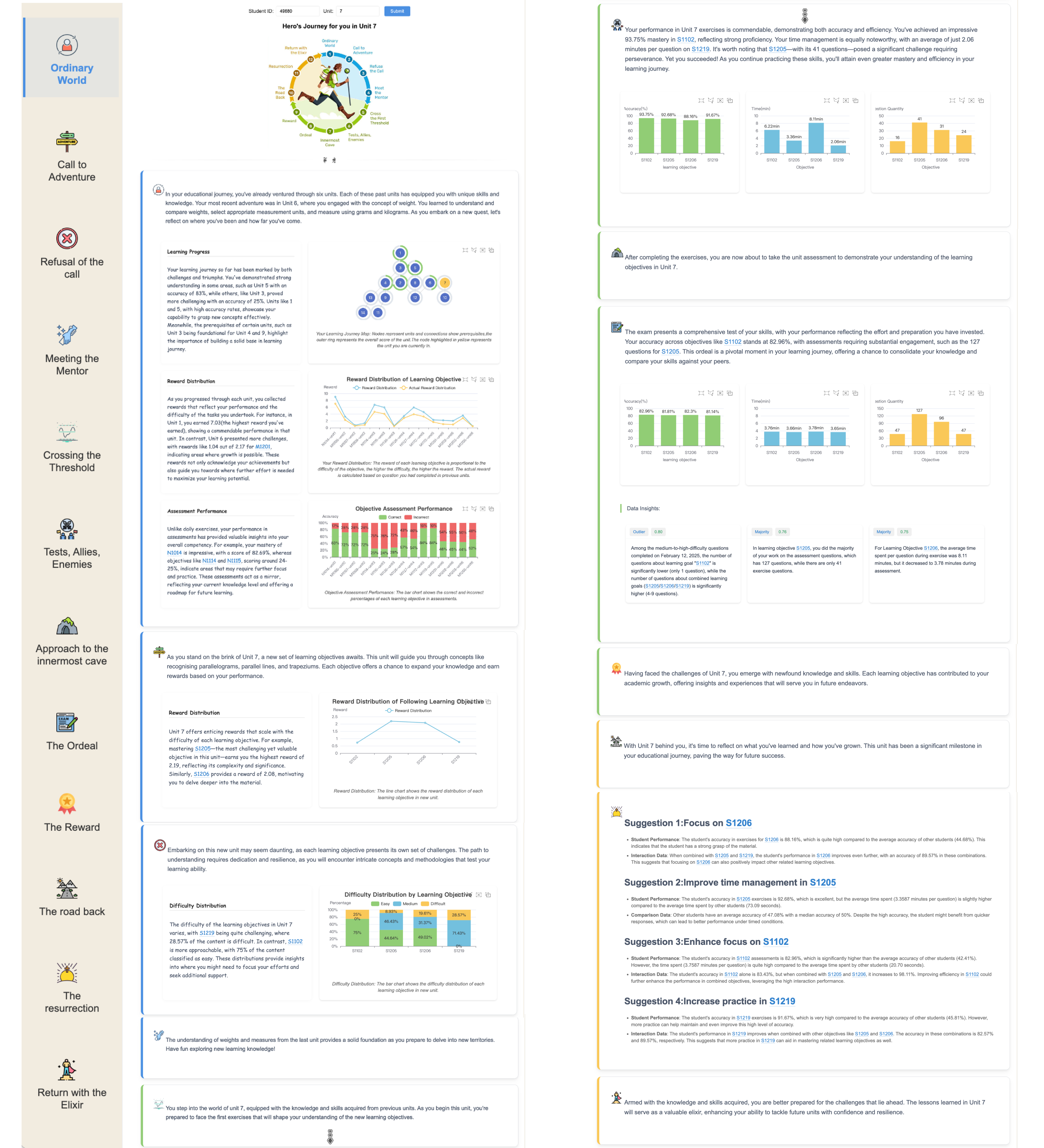}
\vspace{-5px}
\caption{Report sample (as shown in the body text) for a student with good performance in unit 7. The left sidebar serves as an interactive index for rapid navigation, while the right pane displays detailed content.}
\label{fig: report_old}
\vspace{-10px}
\end{figure*}

\begin{figure*}[t]
\centering
\includegraphics[width=0.98\linewidth]{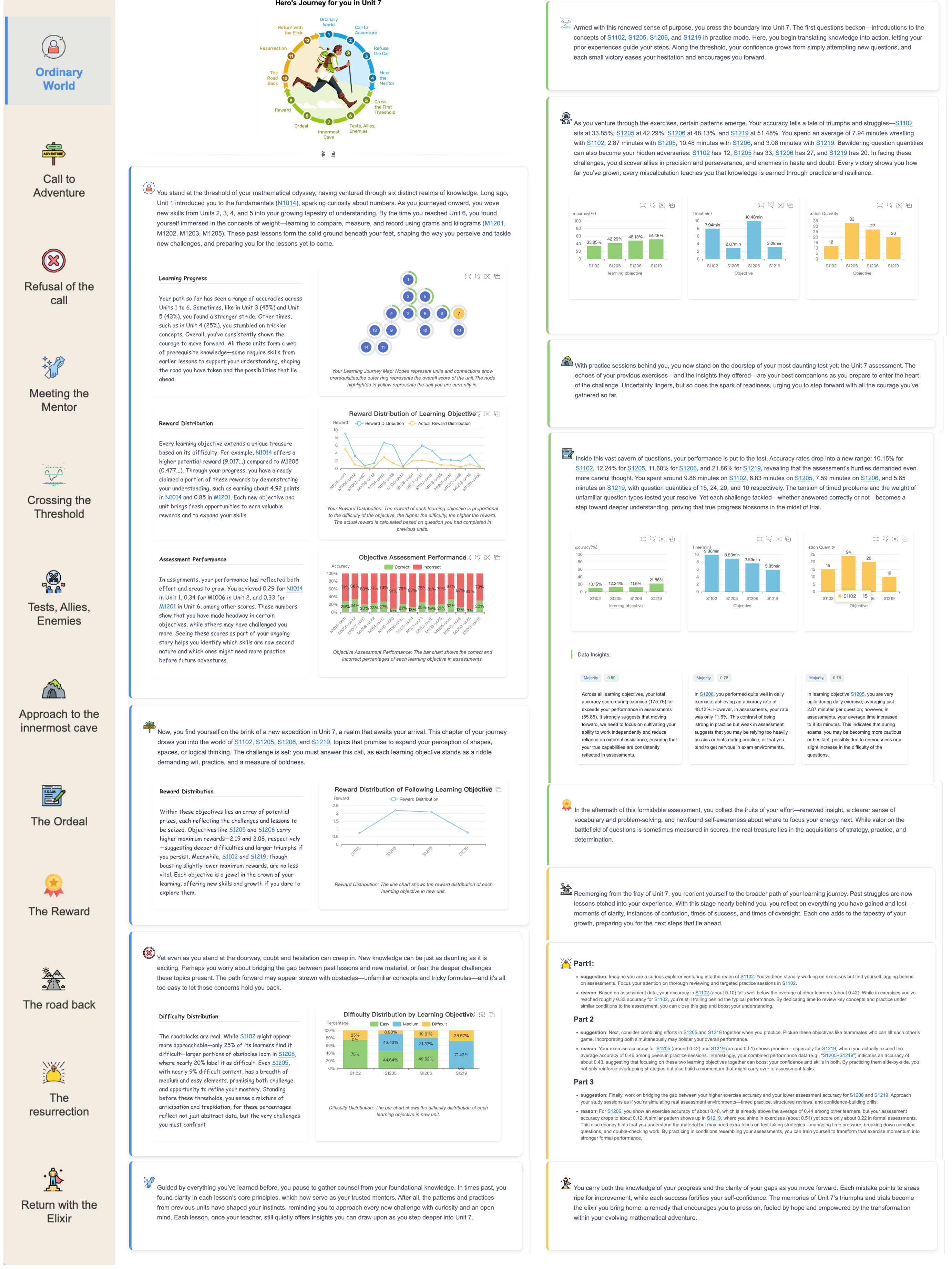}
\vspace{-5px}
\caption{Report sample for a student with lower proficiency. The system provides supportive feedback and tailored recommendations to address identified knowledge gaps.
} 
\label{fig: report2}
\vspace{-10px}
\end{figure*}

\begin{figure*}[t]
\centering
\includegraphics[width=0.98\linewidth]{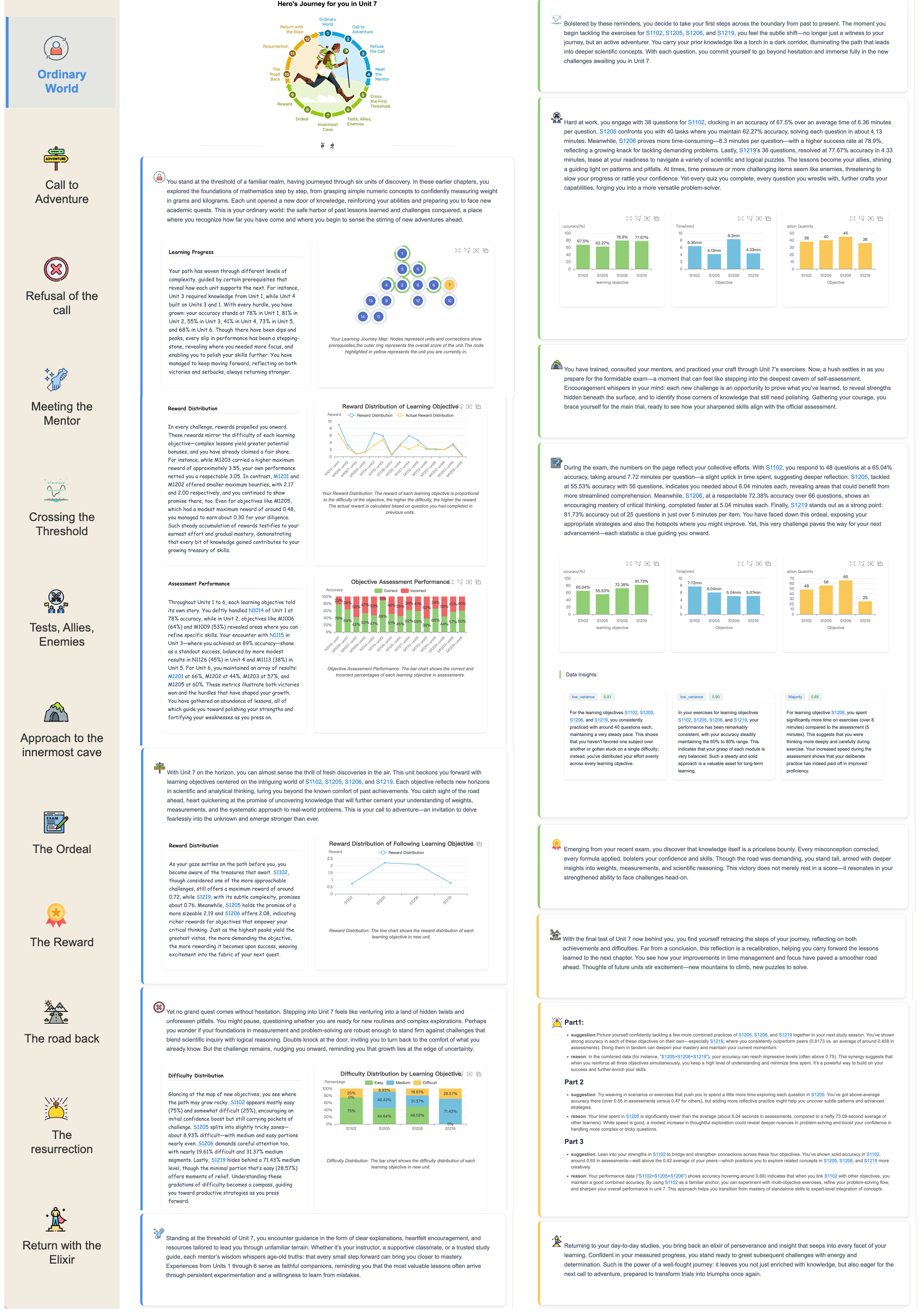}
\vspace{-5px}
\caption{Report sample for a student maintaining a steady learning pace. The system validates their study habits and encourages continued adherence to the current learning plan.
} 
\label{fig: report3}
\vspace{-10px}
\end{figure*}

\begin{figure*}[t]
\centering
\includegraphics[width=0.75\linewidth]{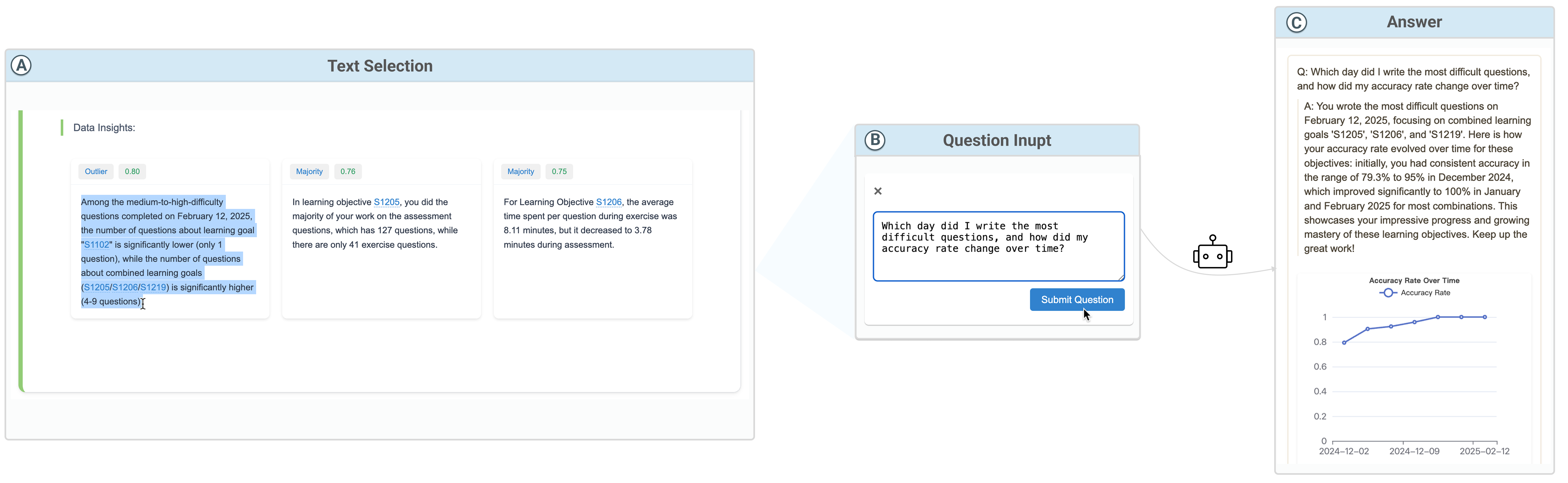}
\vspace{-5px}
\caption{Examples of interaction modules. The user highlights specific learning objectives to query their historical progress. The system responds with a temporal visualization of difficulty and accuracy.
} 
\label{fig: intcase1}
\vspace{-10px}
\end{figure*}

\begin{figure*}[t]
\centering

\includegraphics[width=0.75\linewidth]{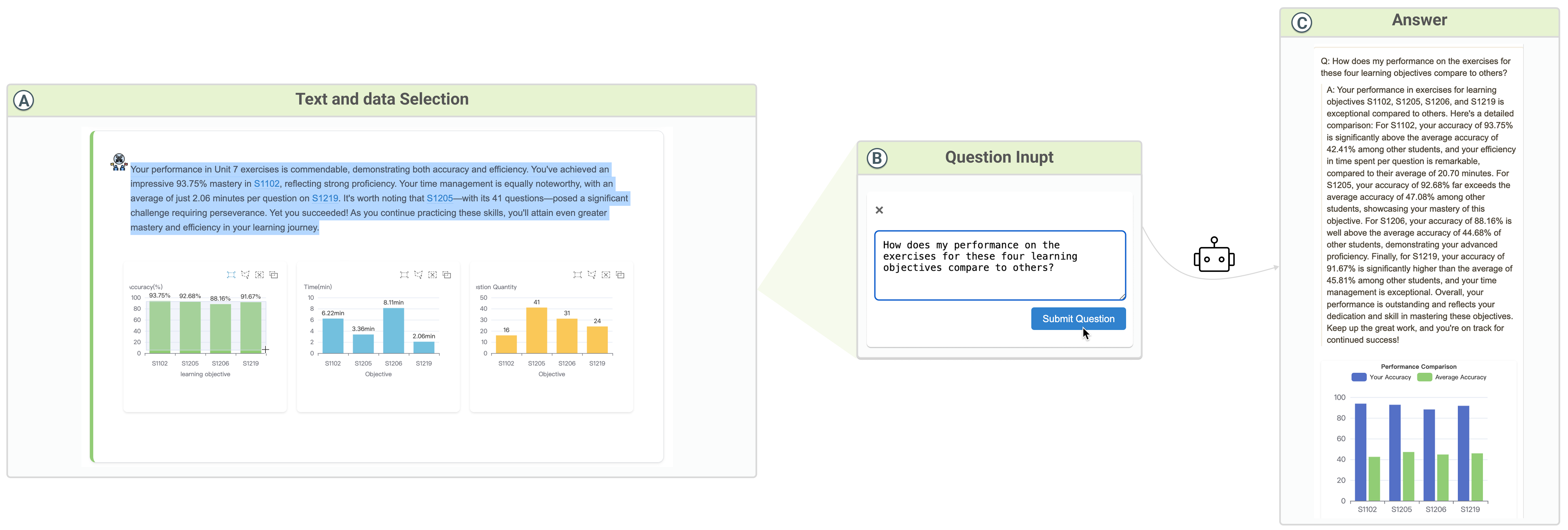}
\vspace{-5px}
\caption{Examples of interaction modules. The user queries how their performance aligns with that of other students. The system contextualizes the answer with a chart comparing individual scores against the cohort average.
} 
\label{fig: intcase2}
\vspace{-10px}
\end{figure*}

\begin{figure*}[t]
\centering

\includegraphics[width=0.75\linewidth]{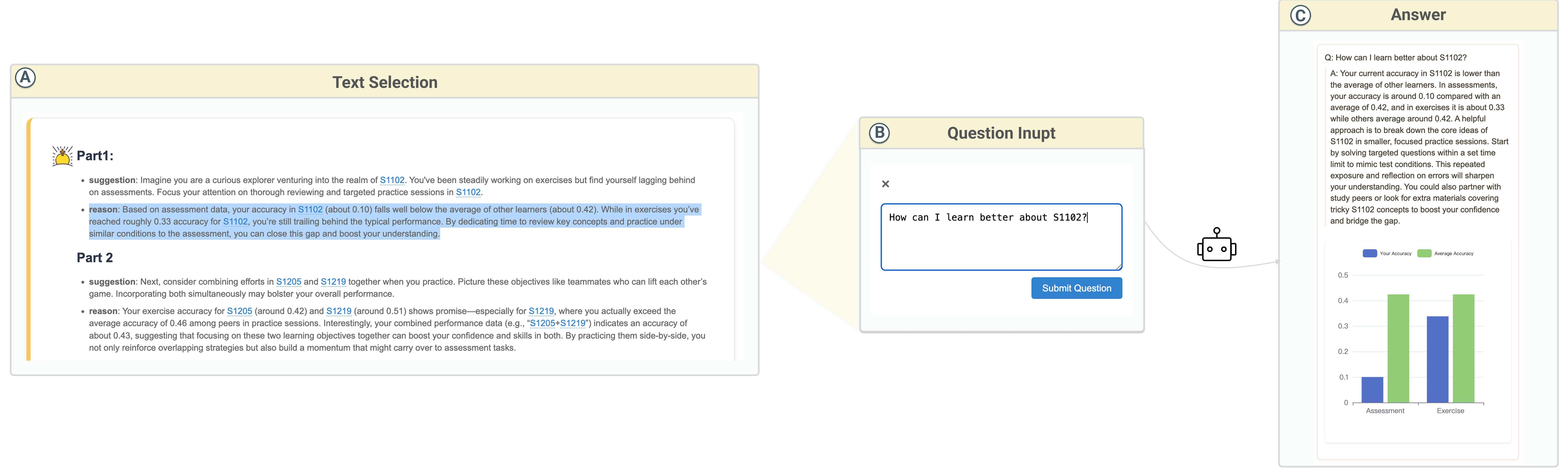}
\vspace{-5px}
\caption{Examples of interaction modules. The user questions a specific learning suggestion for learning objective S1102. The agent visualizes the performance gap that triggered this recommendation, helping the user understand the rationale behind the advice.
} 
\label{fig: intcase3}
\vspace{-10px}
\end{figure*}

\begin{figure*}[t]
\centering
\includegraphics[width=0.75\linewidth]
{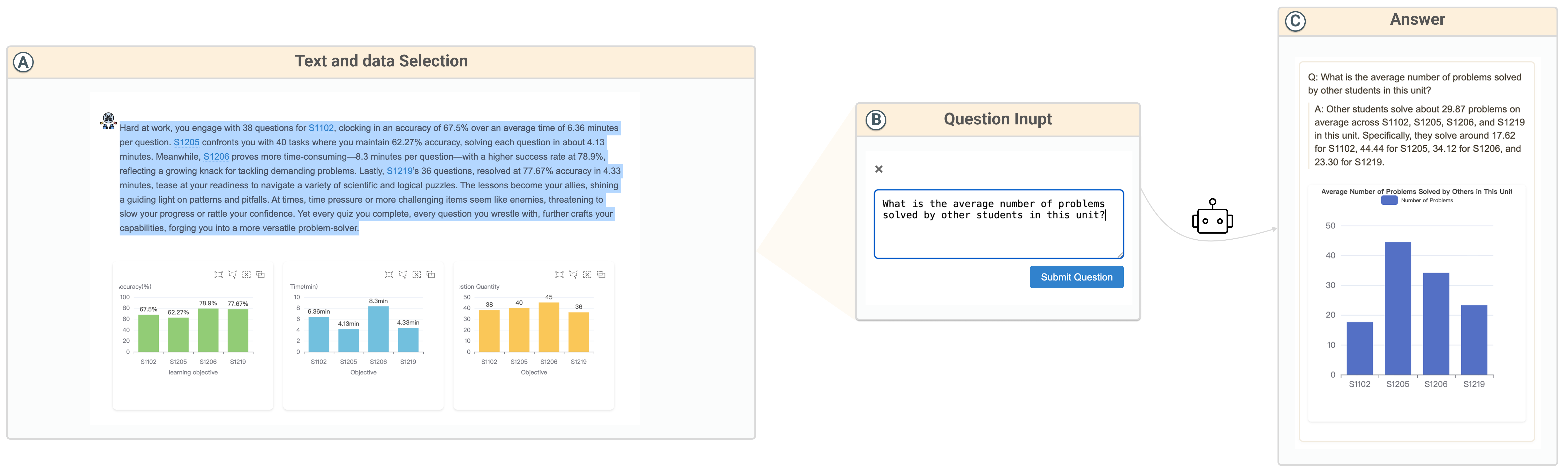}
\vspace{-5px}
\caption{Examples of interaction modules. The user requests specific cohort data (average problems solved) that was not explicitly visualized in the original report. The agent retrieves and plots this data across multiple objectives.
} 
\label{fig: intcase4}
\vspace{-10px}
\end{figure*}

We also provide four examples for reference, shown in Figures~\ref{fig: intcase1} to \ref{fig: intcase4}. Specifically, Figures~\ref{fig: intcase1} and \ref{fig: intcase2} demonstrate the Q\&A for Report Sample One (Figure \ref{fig: report_old}), while Figure~\ref{fig: intcase3} and Figure~\ref{fig: intcase4} correspond to Report Sample Two (Figure \ref{fig: report2}) and Report Sample Three (Figure \ref{fig: report3}), respectively.

\end{document}